\UseRawInputEncoding
\documentclass[aps,prb,twocolumn,superscriptaddress]{revtex4-2}

\usepackage[table]{xcolor}
\usepackage{graphicx}% Include figure files
\usepackage{dcolumn}% Align table columns on decimal point
\usepackage{bm}% bold math
\usepackage{hyperref}% add hypertext capabilities
\begin{document}

\preprint{APS/123-QED}

\title{DPmoire: A tool for constructing accurate machine learning force fields in moir\'e systems}% Force line breaks with \\

\author{Jiaxuan Liu}
\affiliation{%
Beijing National Laboratory for Condensed Matter Physics and Institute of Physics,
Chinese Academy of Sciences, Beijing 100190, China
}
\affiliation{%
University of Chinese Academy of Sciences, Beijing 100049, China
}

\author{Zhong Fang}
\author{Hongming Weng}
\email{hmweng@iphy.ac.cn} 
\affiliation{%
Beijing National Laboratory for Condensed Matter Physics and Institute of Physics,
Chinese Academy of Sciences, Beijing 100190, China
}
\affiliation{%
University of Chinese Academy of Sciences, Beijing 100049, China
}

\affiliation{%
Songshan Lake Materials Laboratory, Dongguan, Guangdong 523808, China
}
\author{Quansheng Wu}
\email{quansheng.wu@iphy.ac.cn} 
\affiliation{%
Beijing National Laboratory for Condensed Matter Physics and Institute of Physics,
Chinese Academy of Sciences, Beijing 100190, China
}
\affiliation{%
University of Chinese Academy of Sciences, Beijing 100049, China
}

\date{\today}% It is always \today, today,

\begin{abstract}
In moir\'e systems, the impact of lattice relaxation on electronic band structures is significant, yet the computational demands of first-principles relaxation are prohibitively high due to the large number of atoms involved. To address this challenge, We introduce a robust methodology for the construction of machine learning potentials specifically tailored for moir\'e structures and present an open-source software package \textit{\textbf{DPmoire}} designed to facilitate this process. Utilizing this package, we have developed machine learning force fields (MLFFs) for MX$_2$ (M = Mo, W; X = S, Se, Te) materials. Our approach not only streamlines the computational process but also ensures accurate replication of the detailed electronic and structural properties typically observed in density functional theory (DFT) relaxations. The MLFFs were rigorously validated against standard DFT results, confirming their efficacy in capturing the complex interplay of atomic interactions within these layered materials. This development not only enhances our ability to explore the physical properties of moir\'e systems with reduced computational overhead but also opens new avenues for the study of relaxation effects and their impact on material properties in two-dimensional layered structures.
\end{abstract}

\maketitle

\section{\label{sec:level1}Introduction}
In recent years, two-dimensional twisted moir\'e structures have captured significant interest due to the diverse physical phenomena they exhibit. By varying the interlayer twist angle, researchers can tune the band structure of these materials, enabling the experimental observation of novel phenomena. For instance, in twisted graphene, when the twist angle reaches the so-called "magic angle," the valence band flattens, prompting electrons to transition from a weakly correlated to a strongly correlated state. This shift gives rise to a host of intriguing behaviors, including unconventional superconductivity, Mott insulating states, and the quantum anomalous Hall effect\cite{cao_unconventional_2018, cao_correlated_2018, xie_fractional_2021, bistritzer_moire_2011, liu_quantum_2019, liu_theories_2021, lu_superconductors_2019, kerelsky_maximized_2019, jiang_charge_2019, serlin_intrinsic_2020, xie_spectroscopic_2019, shen_correlated_2020}. Similar phenomena have also been observed in moir\'e bilayers of transition metal dichalcogenides (TMDs)\cite{cai_signatures_2023, jia_moire_2024, wang_fractional_2023, zeng_thermodynamic_2023, li_quantum_2021, wang_correlated_2020, devakul_magic_2021}.

In twisted structures, the moir\'e potential narrows the bandwidth as the periodicity of the structure increases. For instance, the bandwidth of bilayer twisted graphene at a twist angle of 1.08$^{\circ}$ is only a few meV\cite{bistritzer_moire_2011, leconte_relaxation_2022}, while the bandwidth of bilayer twisted MoTe$_2$ at 3.89$^{\circ}$ is just over 10 meV\cite{jia_moire_2024}. Such narrow bands are highly susceptible to the effects of lattice relaxation, which significantly influences their electronic properties. Theoretical calculations reveal that the electronic band structures of rigid twisted graphene differ markedly from those of relaxed systems\cite{leconte_relaxation_2022}. Additionally, experimental studies using scanning tunneling microscopy (STM) have also documented the relaxation patterns in TMDs resulting from lattice reconstruction\cite{shabani_deep_2021, tilak_moire_2023}.

To accurately model the electronic properties of moir\'e structures, density functional theory (DFT) is often employed, particularly for structures with large twist angles, where it is considered essential for reliable structural relaxation ~\cite{jia_moire_2024, wang_fractional_2023, devakul_magic_2021}. However, despite its high level of accuracy, the computational complexity of DFT scales cubically with the number of atoms, rendering it impractical for smaller-angle structures due to the sheer number of atoms involved.

To address this computational challenge, researchers have developed parameterized continuum models that are better suited for structures with small twist angles ~\cite{carr_relaxation_2018, nakatsuji_multiscale_2023, jung_origin_2015, jung_ab_2014, koshino_effective_2020, nam_lattice_2017, koshino_maximally_2018, miao_truncated_2023}. While these models provide a computationally feasible alternative, they typically do not reach the accuracy levels of DFT relaxation. For materials such as graphene~\cite{stuart_reactive_2000, ouyang_nanoserpents_2018, kolmogorov_registry-dependent_2005} and transition metal dichalcogenides (TMDs)~\cite{naik_kolmogorovcrespi_2019}, empirical force fields have been effectively utilized for structural relaxation~\cite{nielsen_accurate_2023, long_atomistic_2022, herzog-arbeitman_moire_2024, shen_correlated_2020, haddadi_moire_2020}. However, in other systems, robust and extensively validated empirical potentials remain scarce, limiting the scope of studies that can be conducted.

Machine learning force fields (MLFF) offer a promising solution to the computational challenges posed by moir\'e structures\cite{jinnouchi_--fly_2019, musaelian_learning_2022, batzner_e3-equivariant_2022, wang_deepmd-kit_2018, zhang_deep_2018, zeng_deepmd-kit_2023, behler_generalized_2007, bartok_gaussian_2010, schutt_schnet_2018, drautz_atomic_2019, park_accurate_2021, xie_bayesian_2021}. Recent advancements in universal MLFFs have shown great promise in terms of versatility, efficiency, and accuracy for materials discovery and high-throughput calculations \cite{chen2022universal, deng2023chgnet, batatia_foundation_2024, choudhary_unified_2023, chen_graph_2019, xie_gptff_2024}. However, in the context of moir\'e systems, the energy scales of electronic bands are often on the order of millielectron volts (meV), a range comparable to the accuracy limits of these universal MLFFs. This indicates that while universal MLFFs provide broad applicability, their precision may be insufficient for structural relaxation tasks in moir\'e systems, necessitating the development of MLFFs specifically tailored to individual material systems.

Previous efforts have successfully constructed MLFFs for twisted structures, achieving encouraging outcomes. Some studies have developed MLFFs for large twist angles and then applied these models to smaller angles \cite{liu_moire_2022, zhang_polarization-driven_2024}, while others have trained MLFFs on non-twisted structures before using them to relax twisted configurations \cite{jia_moire_2024, zhang_universal_2024}. Additionally, a few approaches have combined initial training on non-twisted structures with subsequent transfer learning on large twist-angle structures to efficiently relax twisted configurations \cite{xu_multiple_2024, mao_transfer_2024}. This multifaceted strategy highlights the adaptability of MLFFs in addressing the specific challenges posed by the diverse configurations encountered in moir\'e systems.

While these innovative approaches have shown promise, their validation has often been limited to specific materials, and a comprehensive tool for constructing MLFFs tailored to twisted structures is still lacking. Moir\'e systems offer a unique platform for exploring novel phenomena such as strong correlations and topological states, with numerous experimental and theoretical advances highlighting their potential. Given the rapid development in this field, there is a pressing need for a universal tool that can conveniently and efficiently construct MLFFs for such complex systems. To bridge this gap, we propose a new methodology and introduce an open-source software, \textit{DPmoire}, designed specifically for moir\'e systems. \textit{DPmoire} leverages non-twisted structures to construct training datasets, facilitating the automated generation of MLFFs tailored to the unique challenges of moir\'e systems. This tool aims to streamline the MLFF construction process, enabling researchers to more effectively study and model the intricate behaviors exhibited by twisted materials.

%%%%%%%%%%%Methods%%%%%%%%%%

%Machine Learning Force Field (MLFF) is a great solution to this problem. 

%\subsection{\label{sec:level2}On-the-fly Machine Learning Force Field}
%The VASP On-the-fly MLFF module can improve the MLFF automatically and simultaneously during a MD simulation[figure]. This module is based on Bayesian linear regression, which could directly estimate the error of its prediction without comparing to ab-initio results. When performing a MD simulation, the module would estimate the error of their predictions. If the estimated error is small, the MLFF predicted results will be apply. When the estimated error is large, the MLFF predicted results will be discarded, and it will perform an ab-initio MD step to get an accurate result. This ab-initio data will be collected to the training dataset and be used to train a new MLFF. This procedure will be repeated along the MD simulation.

%These packages would greatly reduce the calculation cost during a MD simulation by applying MLFF MD steps. Since that, we can fully explore the atomic configuration space to construct a high quality dataset efficiently. However, the VASP On-the-fly MLFF module is lightweighted. It helps accelerate the MD simulation, but the MLFF generated by the algorithm is not accurate enough. The RMSE of forces predicted by VASP MLFF is typically at the magnitude of 0.01 eV/\r{A}

\section{\label{sec:level1}Methods}

\subsection{Moir\'e structures}

\begin{figure}[!htpb]
    \centering
    \includegraphics[width=1\linewidth]{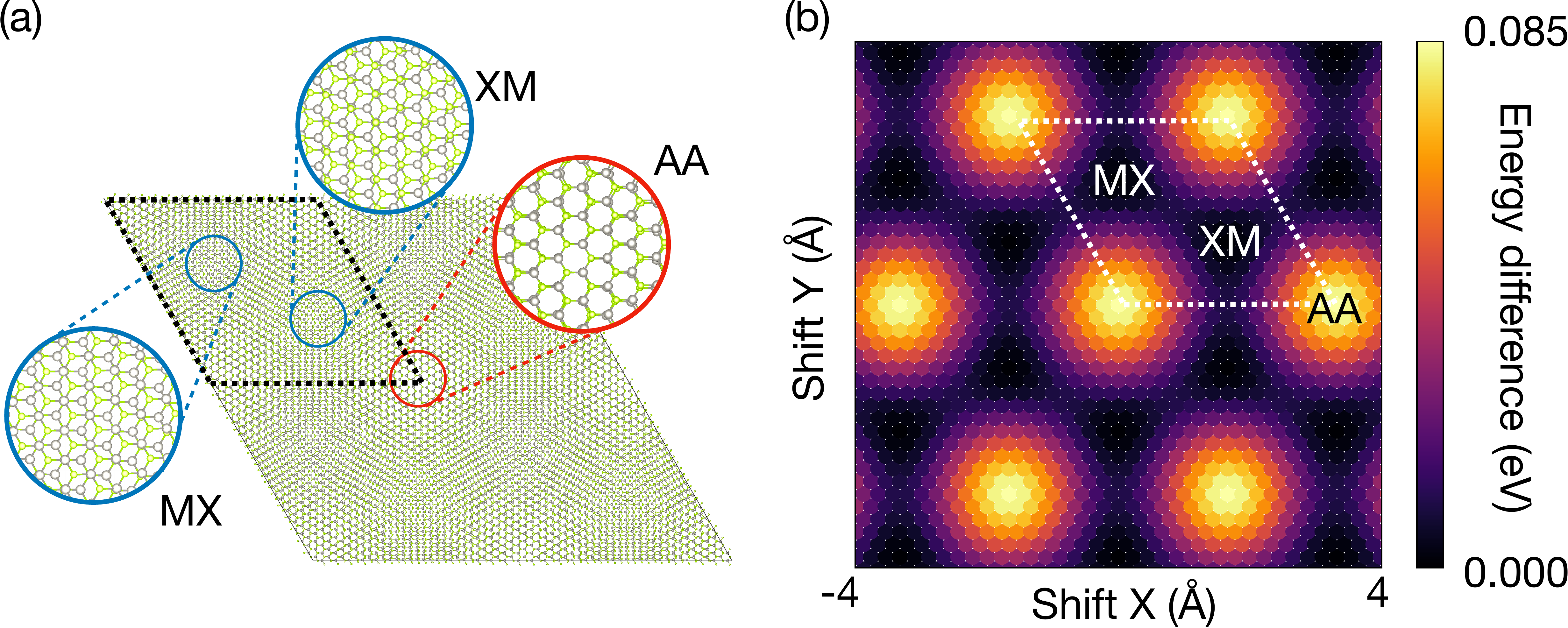}
    \caption{(a) Moir\'e crystal structure of WSe$_2$ with a 2.13$^{\circ}$ AA stacking twist, resembling the atomic layout of non-twisted bilayer WSe$_2$. (b) Energy profile of non-twisted bilayer WSe$_2$ based on relative in-plane shifts between layers, where X and Y axes represent shift vectors, and color indicates unit cell energy. Energy at MX and XM stackings is zeroed. Interlayer distance is 6.8 \AA.}
    \label{fig:moire}
\end{figure}

Moir\'e twisted materials could be constructed by either applying a twist angle between layers of two layered materials or stacking two materials with a slight lattice constant mismatch. Generally, the smaller the twist angle, the larger the resulting moir\'e supercell. Different regions of a moir\'e structure exhibit various stacking arrangements. Taking twisted AA WSe$_2$ as an example (Fig. \ref{fig:moire}), in the AA region, the W/Se atoms in the top layer are aligned with the corresponding W/Se atoms in the bottom layer. In the MX region, the W atoms in the top layer align with the Se atoms in the bottom layer, while in the XM region, the Se atoms in the top layer align with the W atoms in the bottom layer. In non-twisted structures, various stacking configurations correspond to different energy states, as illustrated in Fig. \ref{fig:moire}(b). When the interlayer twist angle is minimal, the lattice vectors of both layers closely match, making the local atomic configurations in the moir\'e structure similar to those in non-twisted structures. By modeling the potential energy surfaces of these non-twisted configurations, we can effectively reconstruct the potential energy landscape of twisted structures, thereby advancing our understanding of their unique properties.

\begin{figure*}
    \centering
    \includegraphics[width=1\linewidth]{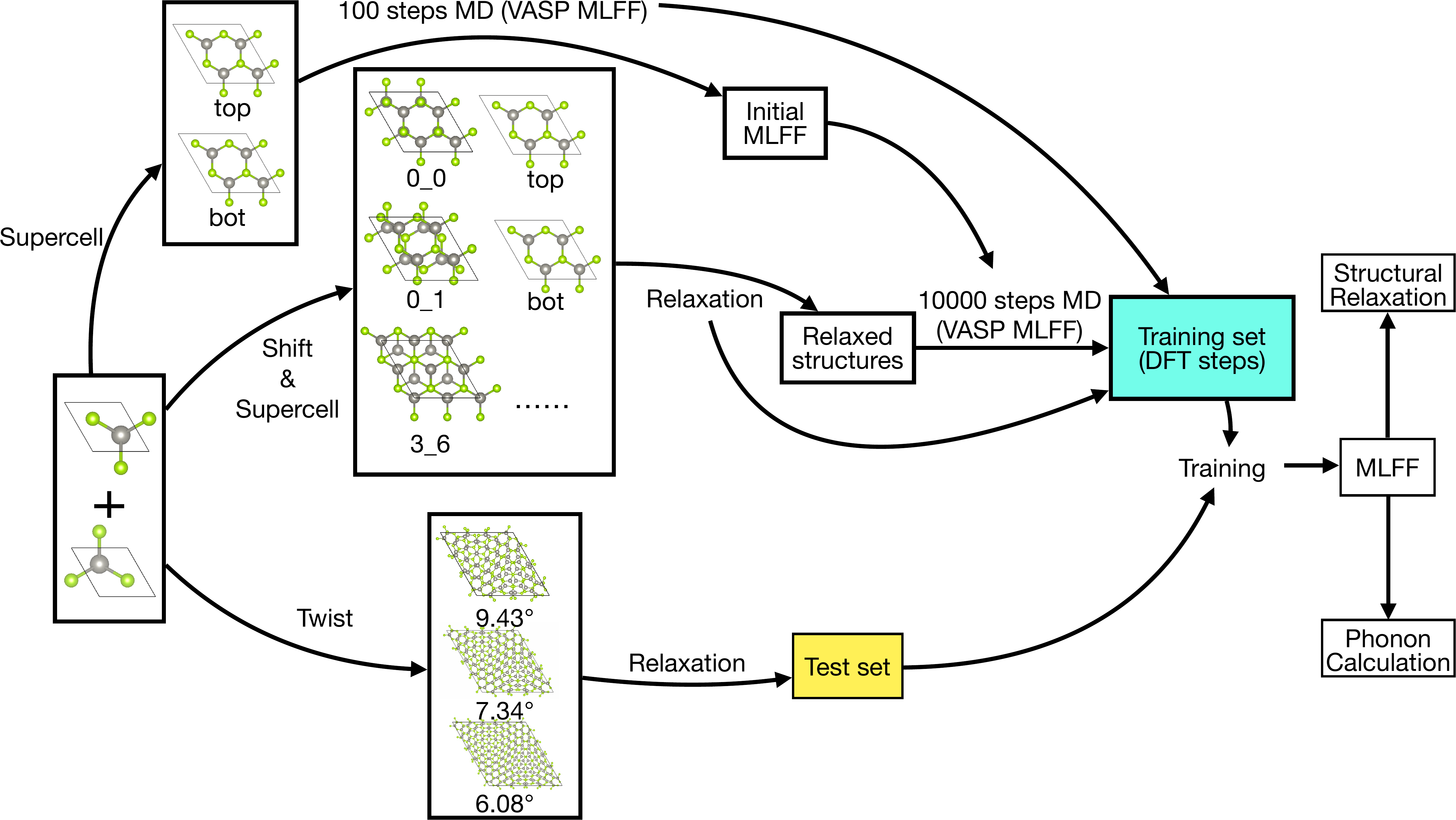}
    \caption{Schematic overview of the process for constructing a Machine Learning Force Field (MLFF) for moir\'e systems. Initially, an MLFF is generated for monolayer structures to stabilize subsequent molecular dynamics (MD) simulations for bilayer systems. We then create non-twisted bilayer structures with various stacking configurations, relax these structures, and run MD simulations using the VASP MLFF module to construct the training dataset. The coordinates (x and y) of a selected atom from each layer are maintained constant during relaxation to preserve the integrity of the stacking order. Subsequently, the twisted structures are relaxed using density functional theory (DFT) to generate the test dataset. The MLFF is ultimately trained on these collected datasets, ensuring it can accurately predict the physical behaviors of moir\'e systems.}
    \label{fig:procedure}
\end{figure*}

\subsection{Machine learning force fields}

Machine learning force field (MLFF)~\cite{jinnouchi_--fly_2019, musaelian_learning_2022, batzner_e3-equivariant_2022, wang_deepmd-kit_2018, zhang_deep_2018, zeng_deepmd-kit_2023, behler_generalized_2007, bartok_gaussian_2010, schutt_schnet_2018, drautz_atomic_2019, park_accurate_2021, xie_bayesian_2021} refers to machine learning algorithms for predicting the energy and forces of crystal structures. Typically, to train an MLFF, it needs a dataset consisting of a set of crystal structures along with their corresponding energies and forces. Once training is complete, the MLFF can rapidly predict the energies and forces of similar structures. The computational cost of MLFF prediction scales linearly with the number of atoms, making the cost of relaxation manageable even for very large structures.

However, constructing a comprehensive dataset can be a time-consuming endeavor. Directly using ab-initio molecular dynamics (MD) simulations to build datasets is a relatively inefficient approach, as structures that are close in time within an MD trajectory are very similar. This similarity results in a redundancy that offers little added value to the training dataset, posing a challenge for efficient MLFF deployment.

On-the-fly machine learning force field (MLFF) approaches like DP-GEN\cite{zhang_dp-gen_2020} and the MLFF module of the Vienna Ab initio Simulation Package (VASP)\cite{jinnouchi_--fly_2019, kresse_efficient_1996} provide effective solutions for managing computational costs in molecular dynamics (MD) simulations. This article focuses on the MLFF module within VASP. This module automates the process of data collection, MLFF training, and its immediate application to accelerate MD simulations within a continuous loop. The MLFF module operates based on Bayesian linear regression, which allows it to directly estimate the error in its predictions without needing to compare them against ab-initio results. During an MD simulation, if the module estimates a small error, it applies the MLFF-predicted results directly. Conversely, if a large error is estimated, it discards these results and performs a density functional theory (DFT) step to obtain accurate data. This ab-initio data is then added to the training dataset for refining the MLFF. This iterative process repeats throughout the MD simulation, allowing for extensive sampling from MD trajectories, which could involve hundreds of thousands of steps, while only requiring DFT calculations for a fraction of those steps. As a result, a high-quality dataset can be constructed with minimal computational expense, optimizing both resources and time.

The MLFF algorithm in VASP is designed to be relatively lightweight, which significantly reduces the training time required during the simulation loop. However, this streamlined approach means that the accuracy of the VASP MLFF may not rival that of more complex neural network-based MLFF algorithms. Consequently, we utilize the VASP MLFF primarily for dataset generation, subsequently employing a more accurate neural network-based MLFF to fit the data collected.

One such advanced approach is NequIP, a machine learning force field based on an E(3)-equivariant graph neural network\cite{batzner_e3-equivariant_2022}. This method ensures covariance among the inputs, outputs, and hidden layers, leading to enhanced data efficiency and model accuracy. Another notable E(3)-equivariant algorithm is Allegro, which is particularly well-suited for large structures and optimized for parallel computing\cite{musaelian_learning_2022}. While this article primarily focuses on the application of Allegro, the dataset generated using our approach is versatile and can be employed to train other MLFF models as well. This flexibility facilitates the exploration and application of various advanced MLFF techniques in computational material science.

\subsection{MLFF for moir\'e systems}
To develop a machine-learned force field (MLFF) for moir\'e superlattice structures, we initially constructed 2 $\times$ 2 supercells of non-twisted bilayers and introduced in-plane shifts to generate various stacking configurations. Subsequently, structural relaxations were performed  for each configuration, ensuring that the x and y coordinates of a reference atom from each layer remained fixed to prevent structural drift toward energetically favorable stackings. The lattice constants were also held constant throughout the simulations. The relaxation data were compiled into a training dataset.

Following the relaxation phase, Molecular Dynamics (MD) simulations were conducted under the aforementioned constraints to augment the training data pool. For these simulations, we employed the VASP MLFF module to explore a wide range of atomic configurations, selectively incorporating data solely from density functional theory (DFT) calculation steps. Given the potential instability when initiating MD simulations with VASP MLFF from an untrained state, we initially established a baseline MLFF using single-layer structures before proceeding with the full simulations. To ensure the MLFF's applicability to moir\'e systems and to mitigate overfitting to non-twisted structures, we constructed the test set using large-angle moir\'e patterns. These were subjected to ab initio relaxations, with the resultant data serving as the test set.

Finally, the compilation of the aforementioned datasets facilitated the training of a robust and accurate MLFF. While we utilized the Allegro framework for MLFF training in this study, other MLFF algorithms, such as DeepMD\cite{wang_deepmd-kit_2018}, could also be effectively trained on these datasets to potentially enhance predictive accuracy and transferability across similar complex structures.

\begin{figure}[!htpb]
    \centering
    \includegraphics[width=1\linewidth]{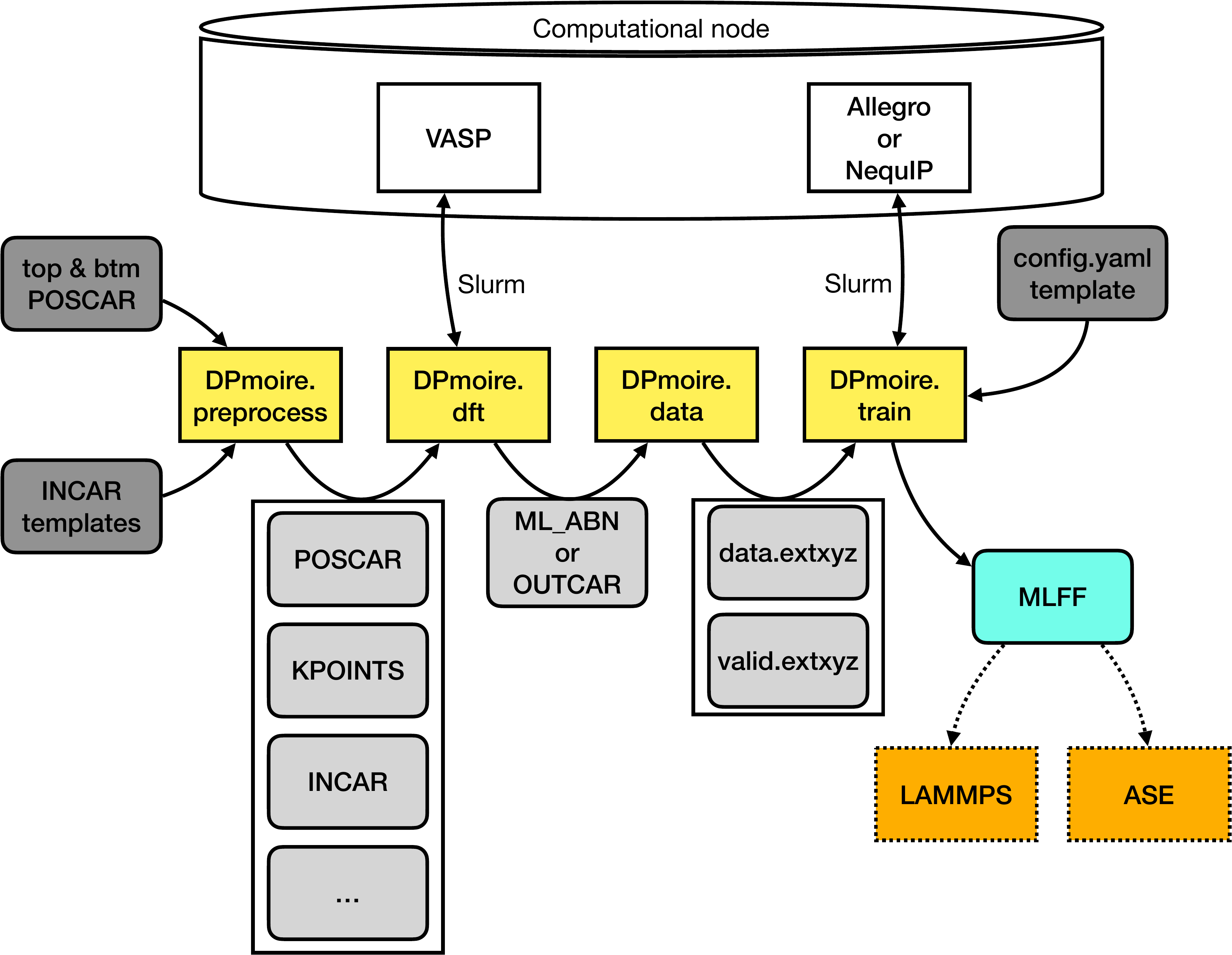}
    \caption{Overview of the DPmoire package workflow. Initially, the \textbf{preprocess} module utilizes the provided POSCAR files for each layer along with an INCAR template to generate the necessary input files for subsequent VASP DFT calculations. The \textbf{dft} module then orchestrates these calculations using the Slurm management system. Upon completion, the \textbf{data} module collects the results and compiles them into datasets formatted in \textbf{extxyz}. Subsequently, the \textbf{train} module begins training a machine learning force field using these datasets, adhering to the parameters specified in the MLFF configuration template file. Once trained, the MLFF can be integrated with software packages such as LAMMPS\cite{LAMMPS} or ASE\cite{larsen2017atomic} to facilitate structural relaxation.}
    \label{fig:DPmoire}
\end{figure}

Eventually, the procedure described above was implemented in DPmoire. As shown in Fig.\ref{fig:DPmoire}, DPmoire is structured into four functional modules: DPmoire.preprocess, DPmoire.dft, DPmoire.data and DPmoire.train. 
Firstly, as provided the unit cell structures of each layer, DPmoire.preprocess module will automatically combine two layers and generate shifted structures of a 2 $\times$ 2 supercell. The twisted structure for building test set will also be prepared. The preprocess module will take care of the input files for VASP according to the provided templates. After that, the DPmoire.dft module will submit VASP calculation jobs through slurm system. When all the calculation is done, the DFT-calculated data in ML\_ABN and OUTCAR files will be collected by DPmoire.data module. Then, DPmoire.data will generate the training set file (data.extxyz) and test set file (valid.extxyz). This format can be directly read by Allegro and NequIP packages. DPmoire.train module will modify the system-dependent settings in configuration file according to given template for training Allegro or NequIP MLFF, and submit the training job. After the training is done, the trained MLFF can be used in ASE\cite{larsen2017atomic} or LAMMPS\cite{LAMMPS} to perform structural relaxation. 

\section{\label{sec:level1}Results}

The accuracy of machine-learned force fields (MLFF) is critically dependent on the precision of underlying density functional theory (DFT) calculations. Particularly in layered materials, the van der Waals (vdW) interactions play a crucial role in determining the DFT-calculated interlayer distances, making their inclusion indispensable. Over the years, a plethora of vdW correction methodologies have been developed \cite{dion_van_2004, klimes_van_2011, klimes_chemical_2010, berland_exchange_2014, sabatini_nonlocal_2013, peng_versatile_2016, ning_workhorse_2022, hamada_van_2014, lee_higher-accuracy_2010, grimme_semiempirical_2006, grimme_consistent_2010, grimme_effect_2011, tkatchenko_accurate_2009, gould_fractionally_2016, gould_c6_2016, kim_universal_2012, steinmann_comprehensive_2011}. Despite these developments, the predicted interlayer distances using different vdW corrections can vary by a few tenths of an \AA ngstrom. 

Given this variation, it is crucial to identify the most appropriate vdW correction for each material prior to the training of MLFFs. To this end, we evaluated the lattice constants obtained under various vdW corrections, comparing them against experimental measurements to ascertain the optimal vdW correction for each material. The details of this comparative analysis are documented in Appendix \ref{appen:vdW corrections}, providing a rigorous foundation for the subsequent MLFF training. Typically, for TMDs, our findings indicate that the DFT-D3 method with a zero-damping function (IVDW=11), as developed by Grimme\cite{grimme_consistent_2010}, provides the optimal vdW correction for MoS$_2$ and WSe$_2$. The vdW-DF-cx method\cite{berland_exchange_2014} proves most effective for MoSe$_2$ and WS$_2$. For MoTe$_2$, the DFT-D2 method (IVDW=10) from Grimme\cite{grimme_semiempirical_2006} yields the best results, while the SCAN+rVV10 correction\cite{peng_versatile_2016} is found to be the optimal vdW correction for WTe$_2$. These tailored corrections are crucial for enhancing the accuracy of DFT calculations, thereby improving the robustness of the developed MLFFs for different TMD materials.

\begin{figure}[!htpb]
    \centering
    \includegraphics[width=1\linewidth]{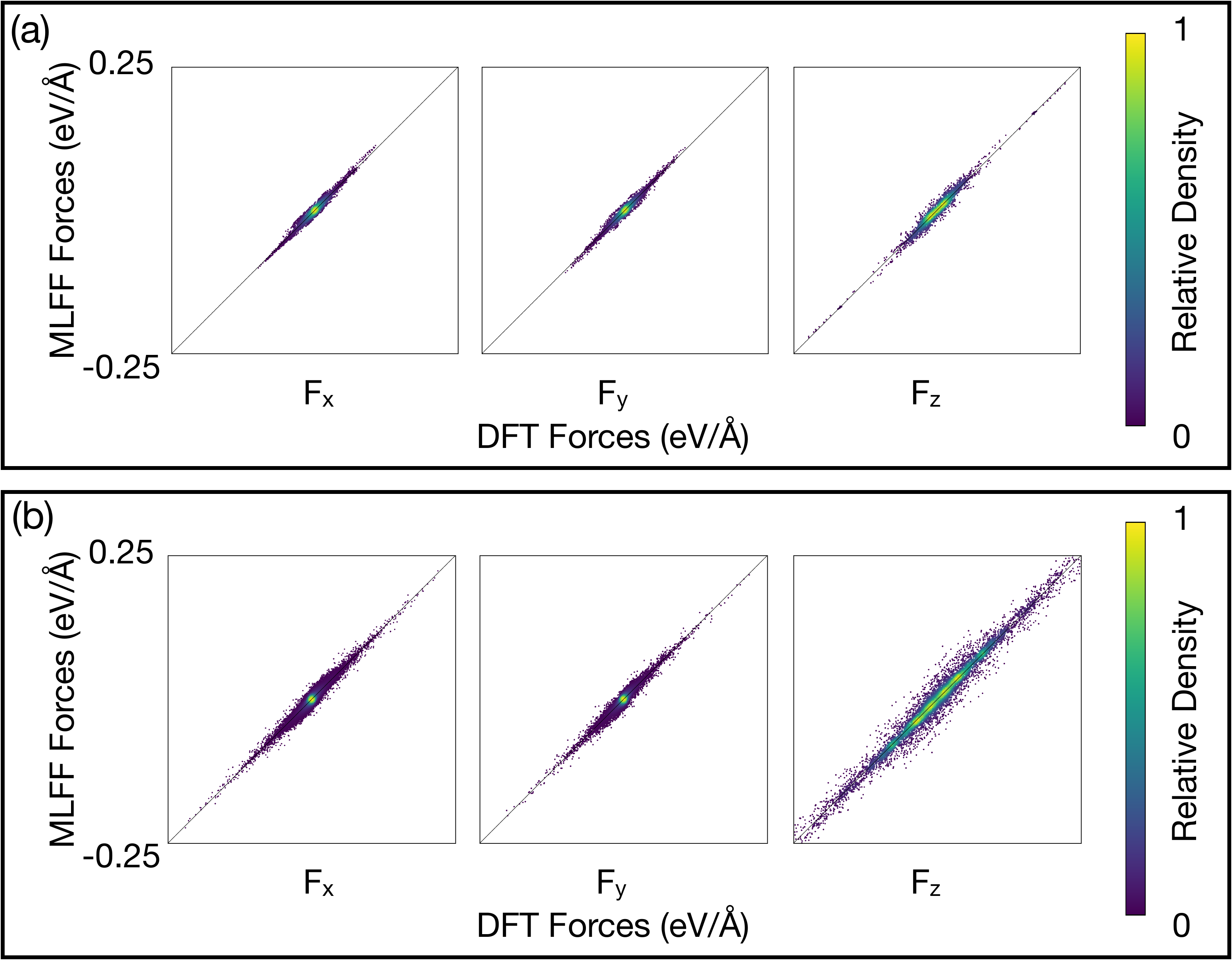}
    \caption{(a) MLFF-predicted versus DFT-calculated forces for AA WSe$_2$ at a 7.34$^{\circ}$ twist. (b) Similar comparison for AA MoS$_2$ across 9.34$^{\circ}$, 7.34$^{\circ}$, and 6.08$^{\circ}$ twists. These panels illustrate the MLFF’s fidelity in capturing force dynamics under different twisting conditions.}
    \label{fig:Ferror}
\end{figure}

Then, the MLFF  is constructed utilizing the previously determined optimal vdW corrections for both AA and AB stacking configurations of MX$_2$ (M = Mo, W; X = S, Se, Te) materials, as thoroughly discussed in Appendix. \ref{appen:MLFF}. We specifically examined AA WSe$_2$ and AA MoS$_2$ as representative examples. The efficacy of the MLFF is demonstrated through a comparison of predicted and DFT-calculated forces within the test set, as illustrated in Fig. \ref{fig:Ferror}. The comparison shows a strong alignment between the MLFF predictions and the DFT calculations, with root mean square errors of 0.007 eV/\AA and 0.014 eV/\AA for WSe$_2$ and MoS$_2$, respectively, underscoring the accuracy of the MLFF in capturing the essential physical interactions in these materials.

\begin{figure*}[!htpb]
    \centering
    \includegraphics[width=1\linewidth]{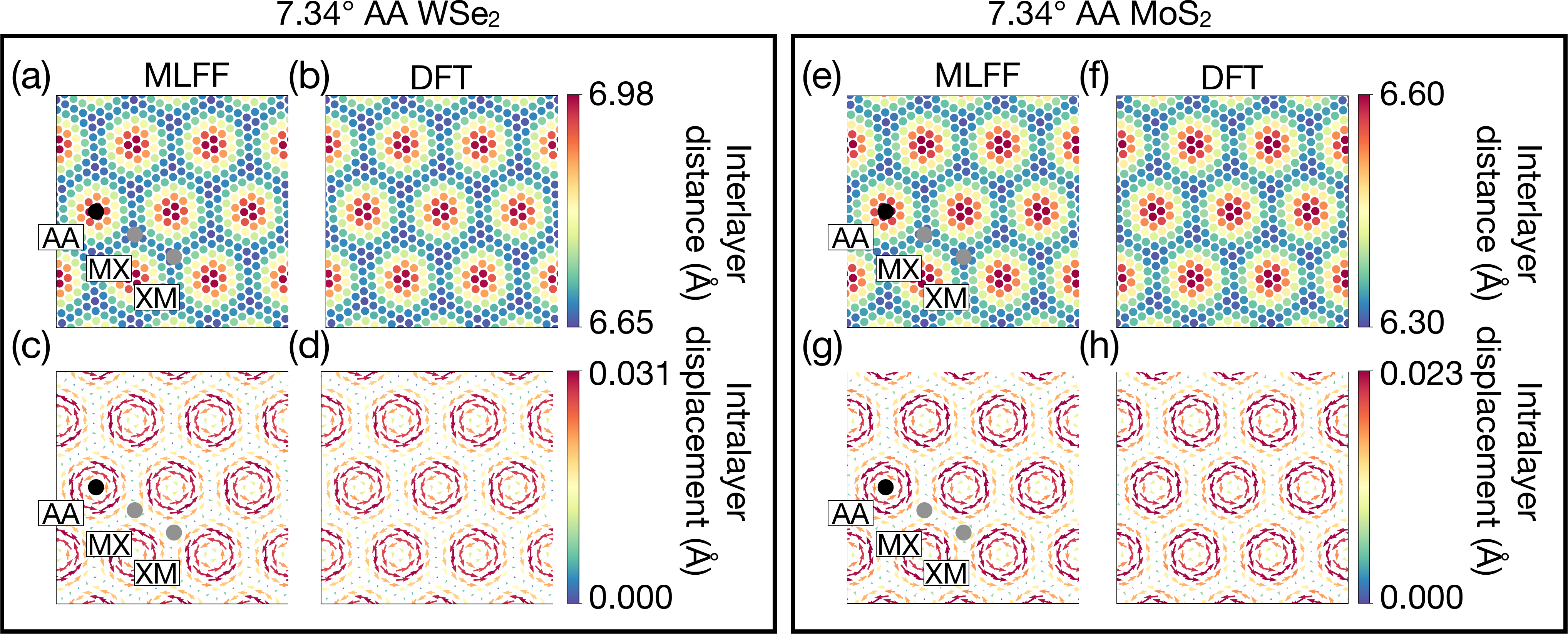}
    \caption{Relaxation pattern of 7.34$^{\circ}$ AA twisted bilayer WSe$_2$. (a) and (b) correspond to the interlayer distance and intralayer displacement in MLFF-relaxed structure, respectively. (c) and (d) cprrespond to the interlayer distance and intralayer displacement in MLFF-relaxed structure, respectively.}
    \label{fig:Relaxation}
\end{figure*}

We further evaluated the performance of the trained MLFFs by relaxing a structure with a 7.34$^{\circ}$ twist angle, followed by a comparison relaxation using DFT. As depicted in Fig.\ref{fig:Relaxation}, the relaxation outcomes from the MLFF are nearly indistinguishable from those obtained via DFT, with no significant deviations observed. The maximum differences in atomic positions were found to be 0.039 \AA in WSe$_2$ and 0.003 \AA in MoS$_2$. In the relaxed structures, regions characterized by MX and XM stacking exhibited lower interlayer distances compared to the AA regions. Throughout the relaxation process, atoms near the AA regions tend to rotate counterclockwise, which intensifies the local twist effect. Conversely, atoms in proximity to the MX and XM regions rotate clockwise. This differential rotation behavior strategically maximizes the area of MX and XM regions while minimizing the AA region. These findings align well with previous theoretical studies~\cite{carr_relaxation_2018}.

\begin{figure*}
    \centering
    \includegraphics[width=1\linewidth]{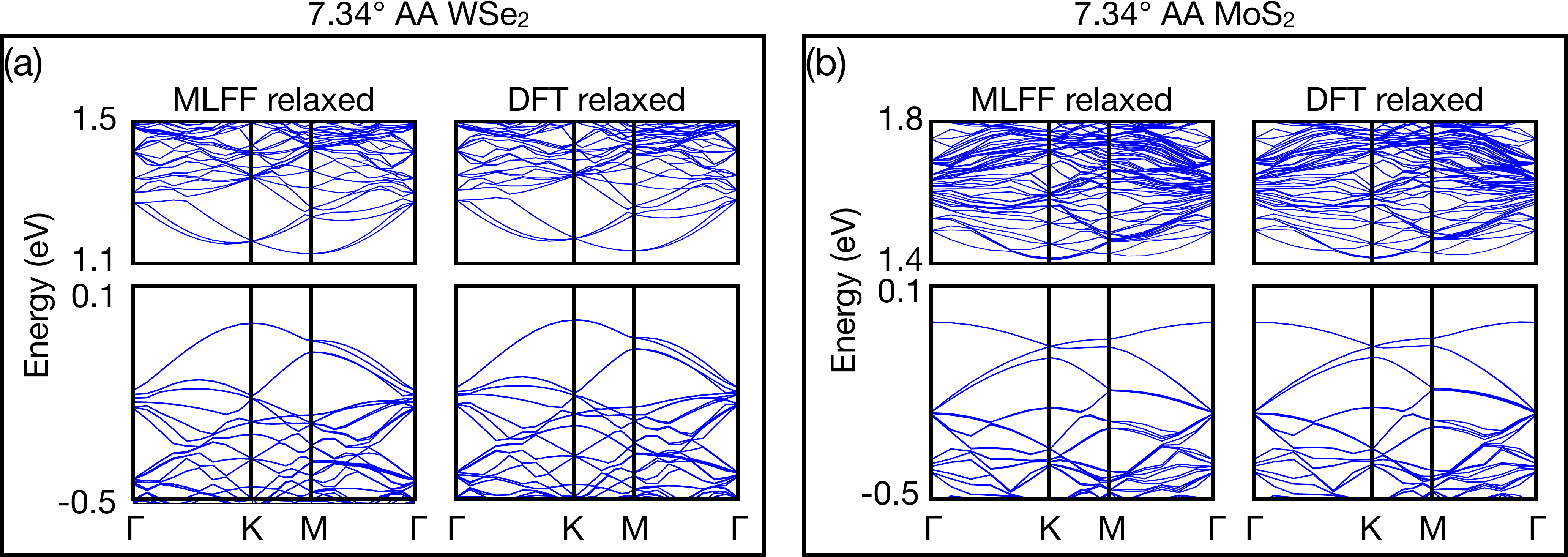}
    \caption{(a), Comparison of electronic band structure between MLFF-relaxed structure and DFT-relaxed structure in 7.34$^{\circ}$ AA WSe$_2$. (b), Comparison of electronic band structure between MLFF-relaxed structure and DFT-relaxed structure in 7.34$^{\circ}$ AA MoS$_2$. }
    \label{fig:bandstructure}
\end{figure*}

We also performed band structure calculations on both MLFF-relaxed and DFT-relaxed structures for AA WSe$_2$ and AA MoS$_2$, as shown in Fig. \ref{fig:bandstructure}. The band structures of the two methods are nearly identical, with only minor differences, demonstrating that the MLFF is sufficiently accurate to capture the essential physical phenomena in moir\'e structures without the need for additional DFT relaxation. As detailed in Appendix \ref{appen:MLFF}, MLFFs for other materials also exhibited robust performance. For MoS$_2$ (Figs. \ref{fig:AA_MoS2}, \ref{fig:AB_MoS2}), WS$_2$ (Figs. \ref{fig:AA_WS2}, \ref{fig:AB_WS2}), AB MoTe$_2$ (Fig. \ref{fig:AB_MoTe2}), and WTe$_2$ (Figs. \ref{fig:AA_WTe2}, \ref{fig:AB_WTe2}), the structures relaxed by MLFF and DFT methods were nearly identical, and their corresponding band structures closely matched. However, for materials like MoSe$_2$ (Figs. \ref{fig:AA_MoSe2}, \ref{fig:AB_MoSe2}) and AA-stacked MoTe$_2$ (Fig. \ref{fig:AA_MoTe2}), slight variations in interlayer distances led to minor differences in their band structures. We further analyzed the 5.09$^{\circ}$ twist angle in AA and AB stacked MoSe$_2$ (Fig. \ref{fig:MoSe2_5.09}), where the discrepancies between DFT-relaxed and MLFF-relaxed structures were reduced, suggesting that the observed suboptimal performance in these materials may be due to the larger twist angles.

\section{\label{sec:level1}Conclusion}
In this work, we introduced a universal methodology and developed an open-source tool, DPmoire, for constructing machine-learned force fields (MLFF) tailored to moir\'e structures. Utilizing the VASP MLFF module, DPmoire effectively generates training sets and constructs validation sets based on large-twist-angle configurations. We successfully trained accurate MLFFs for MX$_2$ (M = Mo, W; X = S, Se, Te) systems, which precisely replicate both the relaxation patterns and electronic band structures observed in DFT relaxations, but at a significantly reduced computational cost.

This innovative tool enables the effective relaxation of moir\'e systems across a broader range of smaller angles and varied materials. Additionally, it facilitates phonon calculations within these complex systems. We anticipate that DPmoire will significantly enhance the understanding of physical phenomena influenced by relaxation effects and spur the discovery of novel moir\'e materials.

\section{\label{sec:level1}Code availability}

The code for DPmoire is publicly available on GitHub at this link \href{https://github.com/JiaxuanLiu-Arsko/DPmoire}{https://github.com/JiaxuanLiu-Arsko/DPmoire}. 
MLFFs for TMDs can be accessed at this release \href{https://github.com/JiaxuanLiu-Arsko/DPmoire/releases/tag/v1.0.0}{https://github.com/JiaxuanLiu-Arsko/DPmoire/releases/tag/v1.0.0}.

\begin{acknowledgments}
J.L. thank Hanqi Pi for helpful discussions. This work was supported by the Science Center of the National Natural Science Foundation of China (Grant No. 12188101), the National Key R\&D Program of China (Grant No. 2023YFA1607400, 2024YFA1408400, 2022YFA1403800), the National Natural Science Foundation of China (Grant No.12274436, 11925408, 11921004), and  H.W. acknowledge support from the New Cornerstone Science Foundation through the XPLORER PRIZE. 
\end{acknowledgments}

\appendix
\renewcommand{\thefigure}{A\arabic{figure}}
\renewcommand{\thetable}{A\arabic{table}}
\setcounter{figure}{0}
\setcounter{table}{0}

\section{van der Waals corrections}
\label{appen:vdW corrections}

For layered materials, the choice of van der Waals correction significantly affects the relaxation results. Therefore, before training the MLFF, we performed DFT relaxation on the bulk structure of each material using different van der Waals corrections. We compared the relaxed lattice constants with the experimental lattice constants from the ICSD \cite{hellenbrandt2004inorganic} to select the optimal lattice constant.

 \begin{table*}
        \begin{tabular}{|c|c|c|c|c|c|}
        \hline
        \textbf{} & \textbf{experiment} & \textbf{optB86\cite{klimes_van_2011}} & \textbf{optB88\cite{klimes_chemical_2010}} & \textbf{vdW-DF\cite{dion_van_2004}} & \textbf{vdW-DF-cx\cite{berland_exchange_2014}} \\
        \hline
        \textbf{a,b(\AA)} & 3.160 & 3.166 & 3.192 & 3.235 & 3.150 \\
        \hline
        \textbf{c(\AA)} & 12.294 & 12.376 & 12.466 & 13.108 & 12.270 \\
        \hline
        \textbf{} & \textbf{optPBE-vdW\cite{klimes_chemical_2010}} & \textbf{rVV10\cite{sabatini_nonlocal_2013}} & \textbf{SCAN+rVV10\cite{peng_versatile_2016}} & \textbf{r$^2$SCAN+rVV10\cite{ning_workhorse_2022}} & \textbf{-} \\
        \hline
        \textbf{a,b(\AA)} & 3.201 & 3.178 & 3.169 & 3.175 & - \\
        \hline
        \textbf{c(\AA)} & 12.717 & 12.130 & 12.456 & 12.427 & - \\
        \hline
        \textbf{} & \textbf{rev-vdW-DF2\cite{hamada_van_2014}} & \textbf{vdW-DF2\cite{lee_higher-accuracy_2010}} & \textbf{IVDW=10} & \cellcolor{lightgray}\textbf{IVDW=11} & \textbf{IVDW=12} \\
        \hline
        \textbf{a,b(\AA)} & 3.167 & 3.284 & 3.188 & \cellcolor{lightgray}3.161 & 3.148 \\
        \hline
        \textbf{c(\AA)} & 12.342 & 12.872 & 12.413 & \cellcolor{lightgray}12.336 & 12.084 \\
        \hline
        \textbf{} & \textbf{IVDW=20} & \textbf{IVDW=21} & \textbf{IVDW=263} & \textbf{IVDW=4} & \textbf{IVDW=3} \\
        \hline
        \textbf{a,b(\AA)} & 3.157 & 3.167 & 3.141 & 3.160 & 3.173 \\
        \hline
        \textbf{c(\AA)} & 12.052 & 12.097 & 12.234 & 12.541 & 12.900 \\
        \hline
        \end{tabular}
     \caption{Relaxaed lattice constant of bulk MoS$_2$ using different van der Waals corrections. Experimental data comes from ICSD 49801. Here IVDW=10 is the DFT-D2 method of Grimme\cite{grimme_semiempirical_2006}. IVDW=11 is the DFT-D3 method of Grimme with zero-damping function\cite{grimme_consistent_2010}. IVDW=12 is the DFT-D3 method with Becke-Johnson damping function \cite{grimme_effect_2011}. IVDW=20 is the Tkatchenko-Scheffler method\cite{tkatchenko_accurate_2009}. IVDW=263 is the Many-body dispersion energy with fractionally ionic model for polarizability method \cite{gould_fractionally_2016, gould_c6_2016}. IVDW=3 is the DFT-ulg method\cite{kim_universal_2012}. IVDW=4 is the dDsC dispersion correction\cite{steinmann_comprehensive_2011}}
     \label{table:vdw_MoS2}
     %experimental result from ICSD 49801
 \end{table*}
 
\begin{table*}
    \begin{tabular}{|c|c|c|c|c|c|}
    \hline
    \textbf{} & \textbf{experiment} & \textbf{optB86} & \textbf{optB88} & \textbf{vdW-DF} & \cellcolor{lightgray}\textbf{vdW-DF-cx} \\
    \hline
    \textbf{a,b(\AA)} & 3.288 & 3.299 & 3.330 & 3.382 & \cellcolor{lightgray}3.279 \\
    \hline
    \textbf{c(\AA)} & 12.92 & 13.044 & 13.161 & 13.881 & \cellcolor{lightgray}12.911 \\
    \hline
    \textbf{} & \textbf{optPBE-vdW} & \textbf{rVV10} & \textbf{SCAN+rVV10} & \textbf{r$^2$SCAN+rVV10} & \textbf{-} \\
    \hline
    \textbf{a,b(\AA)} & 3.341 & 3.316 & 3.295 & 3.308 & - \\
    \hline
    \textbf{c(\AA)} & 13.432 & 12.864 & 13.164 & 13.107 & - \\
    \hline
    \textbf{} & \textbf{rev-vdW-DF2} & \textbf{vdW-DF2} & \textbf{IVDW=10} & \textbf{IVDW=11} & \textbf{IVDW=12} \\
    \hline
    \textbf{a,b(\AA)} & 3.300 & 3.444 & 3.315 & 3.291 & 3.276 \\
    \hline
    \textbf{c(\AA)} & 13.020 & 13.672 & 13.021 & 13.011 & 12.721 \\
    \hline
    \textbf{} & \textbf{IVDW=20} & \textbf{IVDW=21} & \textbf{IVDW=263} & \textbf{IVDW=4} & \textbf{IVDW=3} \\
    \hline
    \textbf{a,b(\AA)} & 3.290 & 3.296 & 3.271 & 3.293 & 3.307 \\
    \hline
    \textbf{c(\AA)} & 12.760 & 12.740 & 12.834 & 13.181 & 13.497 \\
    \hline
    \end{tabular}
    \caption{Relaxaed lattice constant of bulk MoSe$_2$ using different van der Waals corrections. Experimental data comes from ICSD 644335.}
    %experimental result from ICSD 644335
\end{table*}

\begin{table*}[]
\begin{tabular}{|c|c|c|c|c|c|}
\hline
\textbf{} & \textbf{experiment} & \textbf{optB86} & \textbf{optB88} & \textbf{vdW-DF} & \textbf{vdW-DF-cx} \\
\hline
\textbf{a,b(\AA)} & 3.521 & 3.527 & 3.567 & 3.631 & 3.502 \\
\hline
\textbf{c(\AA)} & 13.96 & 14.032 & 14.213 & 15.007 & 13.867 \\
\hline
\textbf{}& \textbf{optPBE-vdW} & \textbf{rVV10} & \textbf{SCAN+rVV10} & \textbf{r$^2$SCAN+rVV10}  & \textbf{PBE} \\
\hline
\textbf{a,b(\AA)} & 3.580 & 3.546 & 3.503 & 3.542 & 3.551 \\
\hline
\textbf{c(\AA)} & 14.475 & 13.949 & 14.223 & 14.187 & 15.095 \\
\hline
\textbf{} & \textbf{rev-vdW-DF2} & \textbf{vdW-DF2} & \cellcolor{lightgray}\textbf{IVDW=10} & \textbf{IVDW=11} & \textbf{IVDW=12} \\
\hline
\textbf{a,b(\AA)} & 3.529 & 3.711 & \cellcolor{lightgray}3.519 & 3.512 & 3.490 \\
\hline
\textbf{c(\AA)} & 14.028 & 14.891 & \cellcolor{lightgray}13.976 & 13.984 & 13.649 \\
\hline
\textbf{} & \textbf{IVDW=20} & \textbf{IVDW=21} & \textbf{IVDW=263} & \textbf{IVDW=4} & \textbf{IVDW=3} \\
\hline
\textbf{a,b(\AA)} & 3.514 & 3.516 & 3.490 & 3.515 & 3.531 \\
\hline
\textbf{c(\AA)} & 13.923 & 13.826 & 13.709 & 14.047 & 14.222 \\
\hline
\end{tabular}
\caption{Relaxaed lattice constant of bulk MoTe$_2$ using different van der Waals corrections. Experimental data comes from ICSD 644476.}
%experimental result from ICSD 644476
\end{table*}

\begin{table*}
\begin{tabular}{|c|c|c|c|c|c|}
\hline
\textbf{} & \textbf{experiment} & \textbf{optB86} & \textbf{optB88} & \textbf{vdW-DF} & \cellcolor{lightgray}\textbf{vdW-DF-cx} \\
\hline
\textbf{a,b(\AA)} & 3.14 & 3.167 & 3.191 & 3.232 & \cellcolor{lightgray}3.152 \\
\hline
\textbf{c(\AA)} & 12.3 & 12.431 & 12.516 & 13.158 & \cellcolor{lightgray}12.332 \\
\hline
\textbf{} & \textbf{optPBE-vdW} & \textbf{rVV10} & \textbf{SCAN+rVV10} & \textbf{r$^2$SCAN+rVV10} & \textbf{-} \\
\hline
\textbf{a,b(\AA)} & 3.200 & 3.179 & 3.155 & 3.167 & - \\
\hline
\textbf{c(\AA)} & 12.777 & 12.186 & 12.509 & 12.482 & - \\
\hline
\textbf{} & \textbf{rev-vdW-DF2} & \textbf{vdW-DF2} & \textbf{IVDW=10} & \textbf{IVDW=11} & \textbf{IVDW=12} \\
\hline
\textbf{a,b(\AA)} & 3.167 & 3.278 & 3.180 & 3.166 & 3.149 \\
\hline
\textbf{c(\AA)} & 12.393 & 12.913 & 12.139 & 12.415 & 12.125 \\
\hline
\textbf{} & \textbf{IVDW=20} & \textbf{IVDW=21} & \textbf{IVDW=263} & \textbf{IVDW=4} & \textbf{IVDW=3} \\
\hline
\textbf{a,b(\AA)} & 3.160 & 3.168 & 3.149 & 3.159 & 3.172 \\
\hline
\textbf{c(\AA)} & 12.232 & 12.223 & 12.464 & 12.584 & 12.979 \\
\hline
\end{tabular}
\caption{Relaxaed lattice constant of bulk WS$_2$ using different van der Waals corrections. Experimental data comes from ICSD 49801.}
%experimental result from ICSD 49801
\end{table*}
\begin{table*}
\begin{tabular}{|c|c|c|c|c|c|}
\hline
\textbf{} & \textbf{experiment} & \textbf{optB86} & \textbf{optB88} & \textbf{vdW-DF} & \textbf{vdW-DF-cx} \\
\hline
\textbf{a,b(\AA)} & 3.285 & 3.298 & 3.327 & 3.377 & 3.279 \\
\hline
\textbf{c(\AA)} & 12.961 & 13.107 & 13.220 & 13.907 & 12.981 \\
\hline
\textbf{} & \textbf{optPBE-vdW} & \textbf{rVV10} & \textbf{SCAN+rVV10} & \textbf{r$^2$SCAN+rVV10} & \textbf{-} \\
\hline
\textbf{a,b(\AA)} & 3.338 & 3.314 & 3.282 & 3.298 & - \\
\hline
\textbf{c(\AA)} & 13.486 & 12.931 & 13.205 & 13.163 & - \\
\hline
\textbf{} & \textbf{rev-vdW-DF2} & \textbf{vdW-DF2} & \textbf{IVDW=10} & \cellcolor{lightgray}\textbf{IVDW=11} & \textbf{IVDW=12} \\
\hline
\textbf{a,b(\AA)} & 3.298 & 3.436 & 3.307 & \cellcolor{lightgray}3.285 & 3.275 \\
\hline
\textbf{c(\AA)} & 13.081 & 13.711 & 12.779 & \cellcolor{lightgray}12.998 & 12.775 \\
\hline
\textbf{} & \textbf{IVDW=20} & \textbf{IVDW=21} & \textbf{IVDW=263} & \textbf{IVDW=4} & \textbf{IVDW=3} \\
\hline
\textbf{a,b(\AA)} & 3.291 & 3.296 & 3.276 & 3.289 & 3.305 \\
\hline
\textbf{c(\AA)} & 12.956 & 12.886 & 13.058 & 13.224 & 13.568 \\
\hline
\end{tabular}
\caption{Relaxaed lattice constant of bulk WSe$_2$ using different van der Waals corrections. Experimental data comes from ICSD 652167.}
%experimental result from ICSD 652167
\end{table*}

\begin{table*}
\begin{tabular}{|c|c|c|c|c|c|}
\hline
\textbf{} & \textbf{experiment} & \textbf{optB86} & \textbf{optB88} & \textbf{vdW-DF} & \textbf{vdW-DF-cx} \\
\hline
\textbf{a,b(\AA)} & 3.491 & 3.530 & 3.570 & 3.632 & 3.505 \\
\hline
\textbf{c(\AA)} & 14.31 & 14.115 & 14.278 & 15.042 & 13.950 \\
\hline
\textbf{} & \textbf{optPBE-vdW} & \textbf{rVV10} & \cellcolor{lightgray}\textbf{SCAN+rVV10} & \textbf{r$^2$SCAN+rVV10} & \textbf{-} \\
\hline
\textbf{a,b(\AA)} & 3.582 & 3.547 & \cellcolor{lightgray}3.498 & 3.539 & - \\
\hline
\textbf{c(\AA)} & 14.538 & 14.019 & \cellcolor{lightgray}14.268 & 14.234 & - \\
\hline
\textbf{} & \textbf{rev-vdW-DF2} & \textbf{vdW-DF2} & \textbf{IVDW=10} & \textbf{IVDW=11} & \textbf{IVDW=12} \\
\hline
\textbf{a,b(\AA)} & 3.531 & 3.711 & 3.552 & 3.510 & 3.494 \\
\hline
\textbf{c(\AA)} & 14.105 & 14.904 & 13.788 & 14.003 & 13.716 \\
\hline
\textbf{} & \textbf{IVDW=20} & \textbf{IVDW=21} & \textbf{IVDW=263} & \textbf{IVDW=4} & \textbf{IVDW=3} \\
\hline
\textbf{a,b(\AA)} & 3.519 & 3.522 & 3.497 & 3.516 & 3.534 \\
\hline
\textbf{c(\AA)} & 14.149 & 14.013 & 13.911 & 14.129 & 14.324 \\
\hline
\end{tabular}

\caption{Relaxaed lattice constant of bulk WTe$_2$ using different van der Waals corrections. Experimental data comes from ICSD 653170.} %exp data from ICSD 653170
\end{table*}

\section{MLFF of transition metal dichalcogenides}
\label{appen:MLFF}
We used DPmoire to construct MLFFs for bilayer twisted MX$_2$ (M = Mo, W; X = S, Se, Te), and the results are presented in this section.

We constructed MLFFs for AA and AB stacking of different materials. To verify the reliability of the MLFFs, we compared the forces calculated by the MLFFs with those from DFT on a test set. Additionally, we constructed a 7.34$^{\circ}$ twisted structure and performed relaxation using both the MLFF and DFT, comparing the interlayer distance and intralayer displacement after relaxation by the different methods. Finally, we calculated the band structures of the structures relaxed by both methods. 

%All materials performed well on the test set. For materials like MoS$_2$ and WS$_2$, the structures relaxed by different methods were very similar, with almost no significant differences, and their band structures also matched very well. For materials such as MoSe$_2$ (Fig. \ref{fig:AA_MoSe2}, \ref{fig:AB_MoSe2}) and AA-stacked MoTe$_2$(Fig. \ref{fig:AA_MoTe2}), there were slight differences in the interlayer distance of the structures, leading to differences in the band structures. We found that the differences in the band structures after relaxation by the two methods were mainly due to the Gamma valley, which is sensitive to the interlayer distance. 

%Even for these materials, the interlayer distance difference between the MLFF-relaxed structures and DFT-relaxed structures was approximately on $10^{-2}$ \r{A} scale, which is smaller than the difference in interlayer distance caused by different choices of van der Waals functionals (as shown in Appendix. \ref{appen:vdW corrections}), and roughly comparable to the difference between the optimal van der Waals functional and experimental values. 

%Furthermore, we relaxed 5.09$^{\circ}$ MoSe$_2$ structures. As shown in Fig. \ref{fig:MoSe2_5.09}, MLFFs perform better in 5.09$^{\circ}$ structures. This indicates that the suboptimal performance of these materials at 7.34$^{\circ}$ is simply due to the large twist angle, which does not satisfy our initial assumption about the similarity between small-angle structures and non-twisted configurations.

\begin{figure}
    \centering
    \includegraphics[width=1\linewidth]{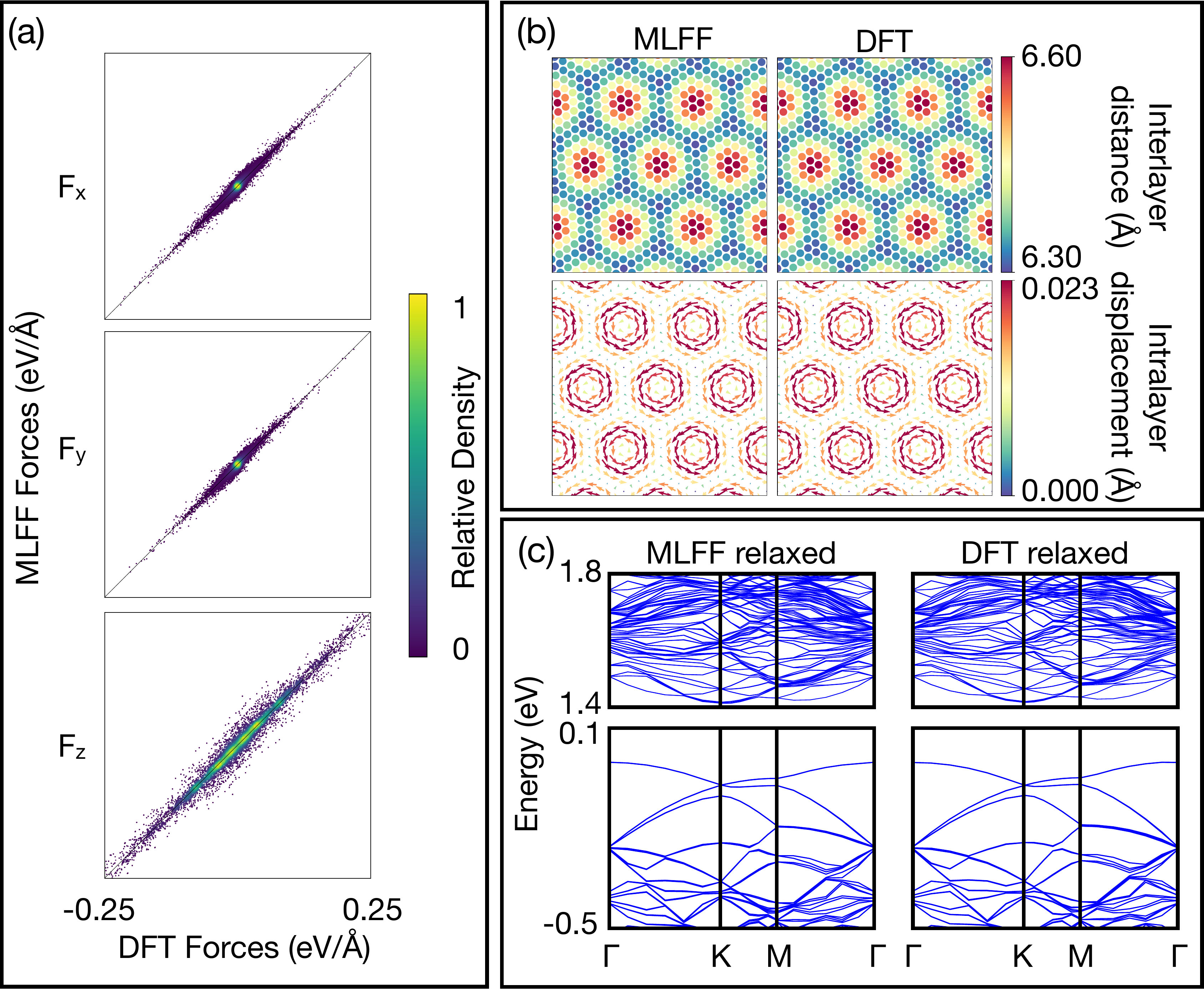}
    \caption{Evaluation of 7.34$^{\circ}$ AA MoS$_2$. (a), Comparison between MLFF-predicted forces and DFT-calculated forces in test set. The x-axis shows the DFT-calculated force, and the y-axis shows the MLFF-calculated force. (b), Relaxation pattern of MLFF-relaxed structure and DFT-relaxed structure. Figs in first row are MLFF-relaxed structure; Figs in second row are DFT-relaxed structure.
    (c), Band structure of MLFF-relaxed structure and DFT-relaxed structure.}
    \label{fig:AA_MoS2}
\end{figure}

 \begin{figure}
    \centering
    \includegraphics[width=1\linewidth]{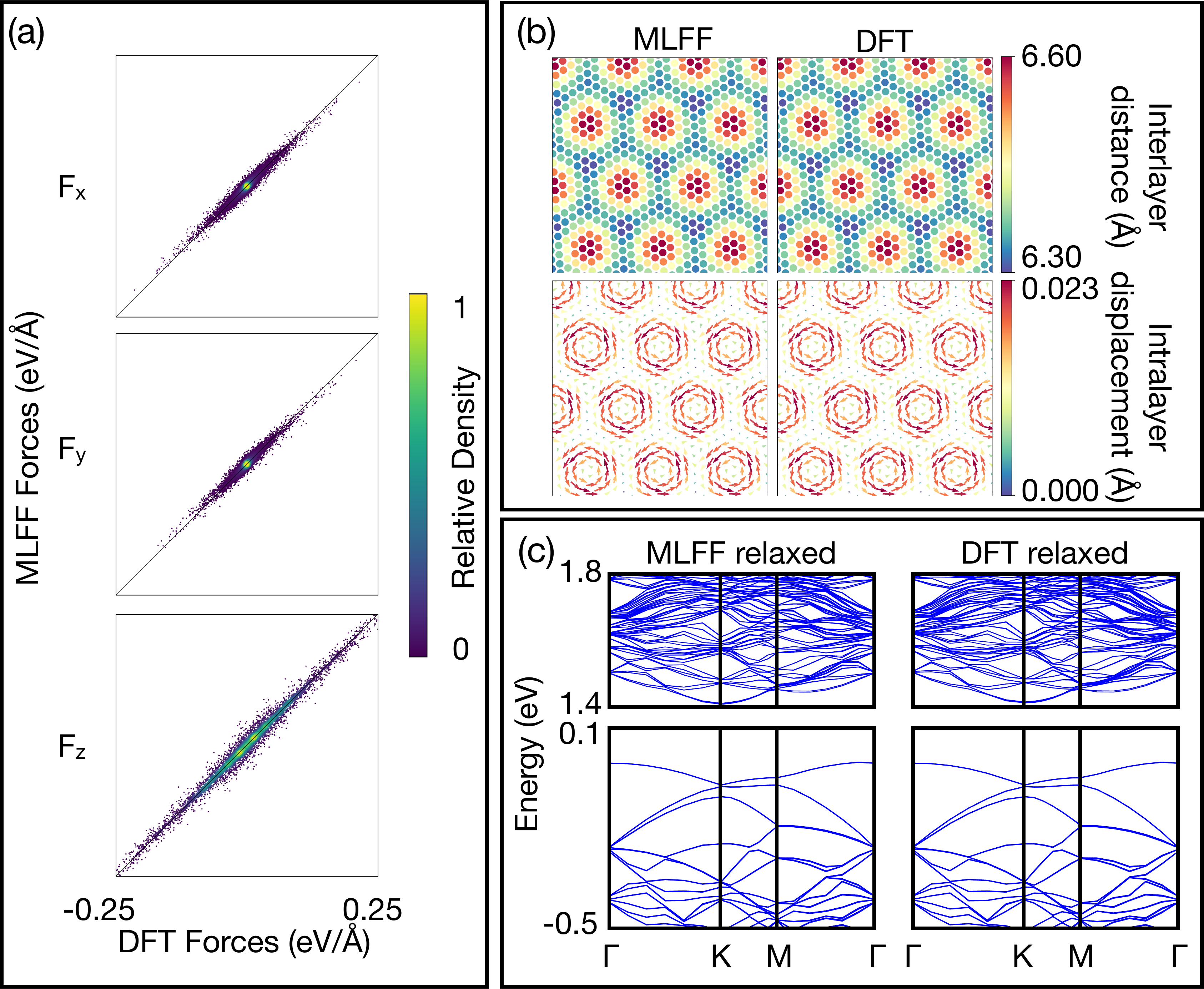}
    \caption{Evaluation of 7.34$^{\circ}$ AB MoS$_2$. (a), Comparison between MLFF-predicted forces and DFT-calculated forces in test set. The x-axis shows the DFT-calculated force, and the y-axis shows the MLFF-calculated force. (b), Relaxation pattern of MLFF-relaxed structure and DFT-relaxed structure. Figs in first row are MLFF-relaxed structure; Figs in second row are DFT-relaxed structure.
    (c), Band structure of MLFF-relaxed structure and DFT-relaxed structure.}
    \label{fig:AB_MoS2}
\end{figure}

\begin{figure}
    \centering
    \includegraphics[width=1\linewidth]{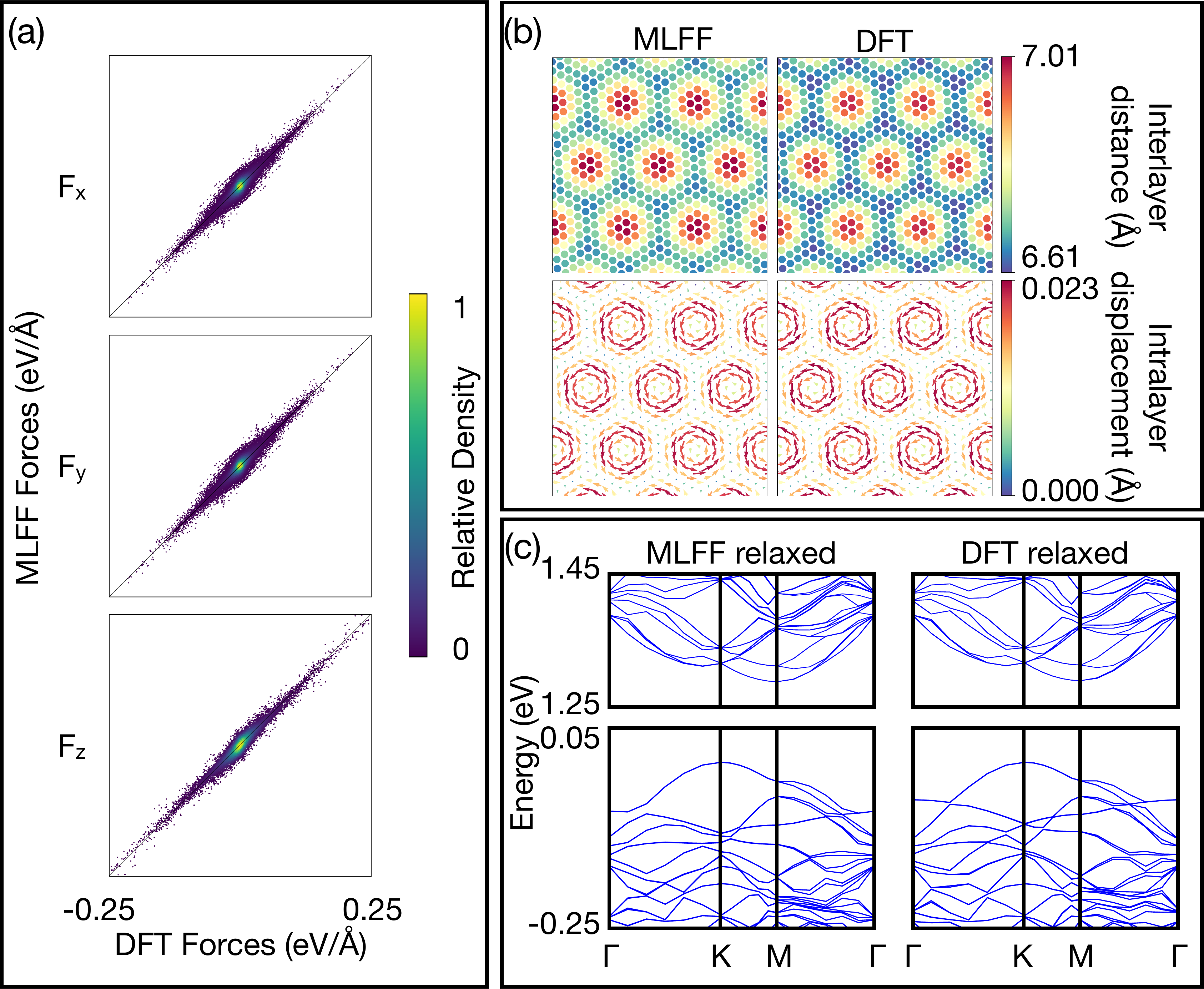}
    \caption{Evaluation of 7.34$^{\circ}$ AA MoSe$_2$. (a), Comparison between MLFF-predicted forces and DFT-calculated forces in test set. The x-axis shows the DFT-calculated force, and the y-axis shows the MLFF-calculated force. (b), Relaxation pattern of MLFF-relaxed structure and DFT-relaxed structure. Figs in first row are MLFF-relaxed structure; Figs in second row are DFT-relaxed structure.
    (c), Band structure of MLFF-relaxed structure and DFT-relaxed structure.}
    \label{fig:AA_MoSe2}
\end{figure}

 \begin{figure}
    \centering
    \includegraphics[width=1\linewidth]{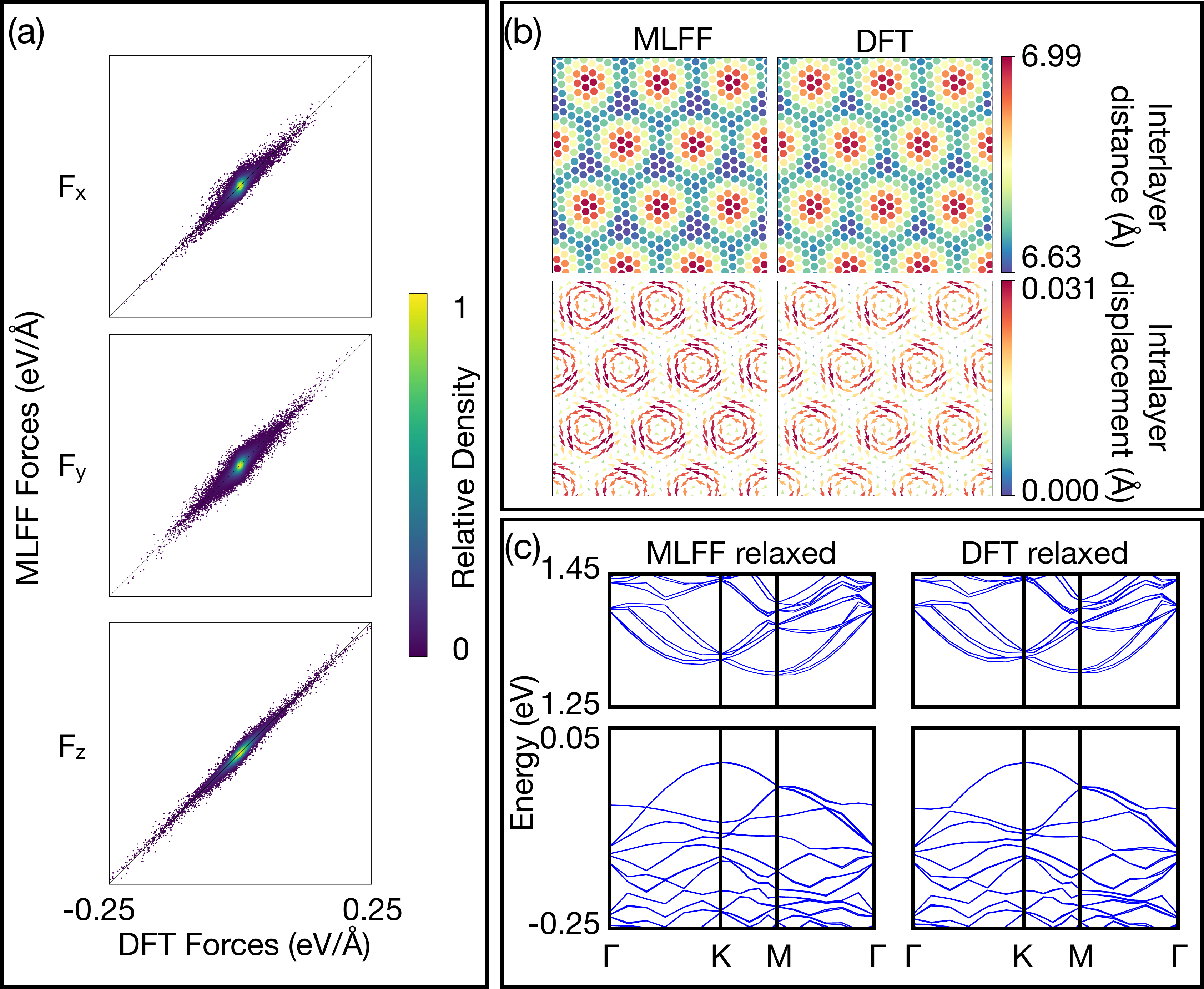}
    \caption{Evaluation of 7.34$^{\circ}$ AB MoSe$_2$. (a), Comparison between MLFF-predicted forces and DFT-calculated forces in test set. The x-axis shows the DFT-calculated force, and the y-axis shows the MLFF-calculated force. (b), Relaxation pattern of MLFF-relaxed structure and DFT-relaxed structure. Figs in first row are MLFF-relaxed structure; Figs in second row are DFT-relaxed structure.
    (c), Band structure of MLFF-relaxed structure and DFT-relaxed structure.}
    \label{fig:AB_MoSe2}
\end{figure}

\begin{figure}
    \centering
    \includegraphics[width=1\linewidth]{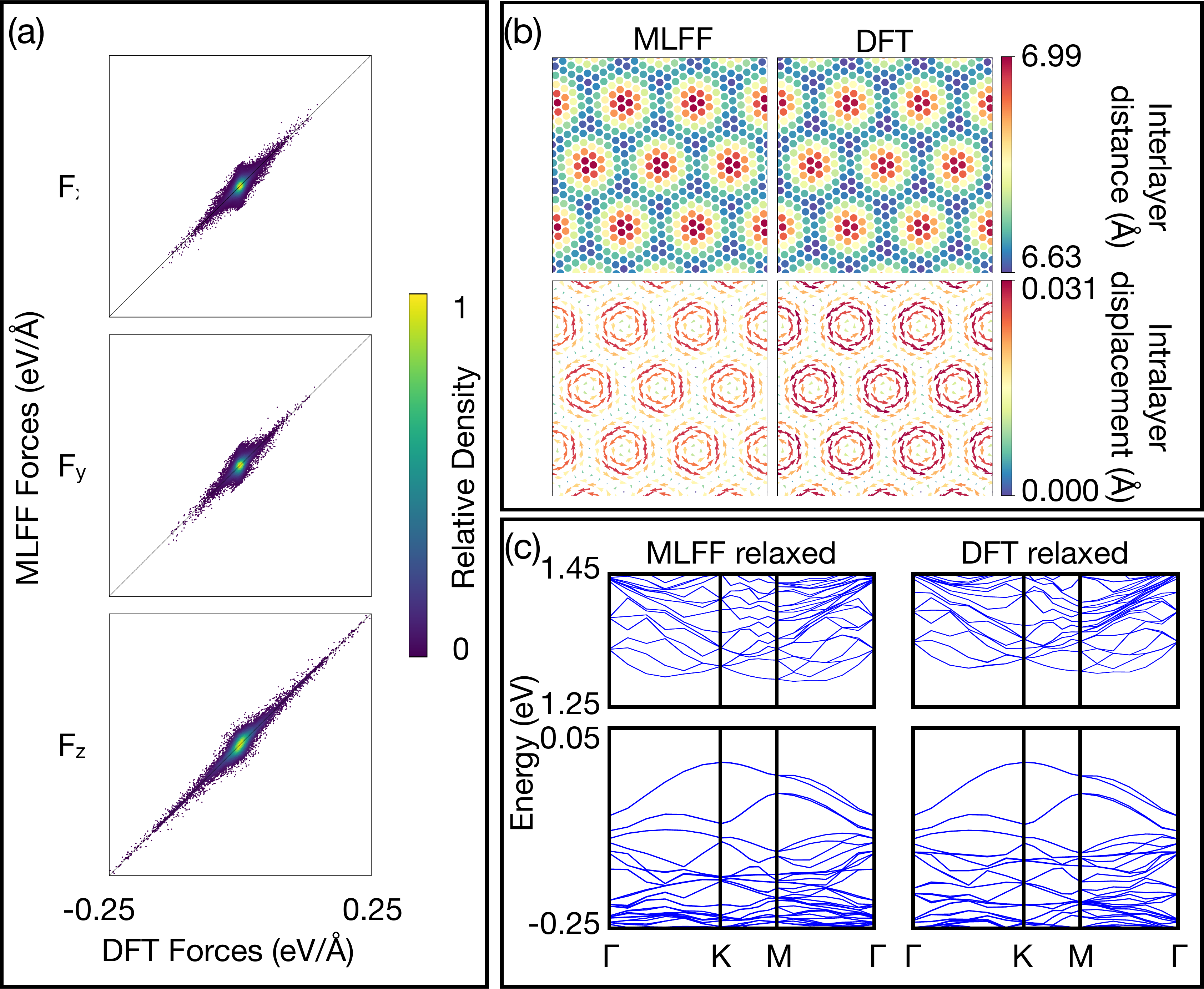}
    \caption{Evaluation of 7.34$^{\circ}$ AA MoTe$_2$. (a), Comparison between MLFF-predicted forces and DFT-calculated forces in test set. The x-axis shows the DFT-calculated force, and the y-axis shows the MLFF-calculated force. (b), Relaxation pattern of MLFF-relaxed structure and DFT-relaxed structure. Figs in first row are MLFF-relaxed structure; Figs in second row are DFT-relaxed structure.
    (c), Band structure of MLFF-relaxed structure and DFT-relaxed structure.}
    \label{fig:AA_MoTe2}
\end{figure}

 \begin{figure}
    \centering
    \includegraphics[width=1\linewidth]{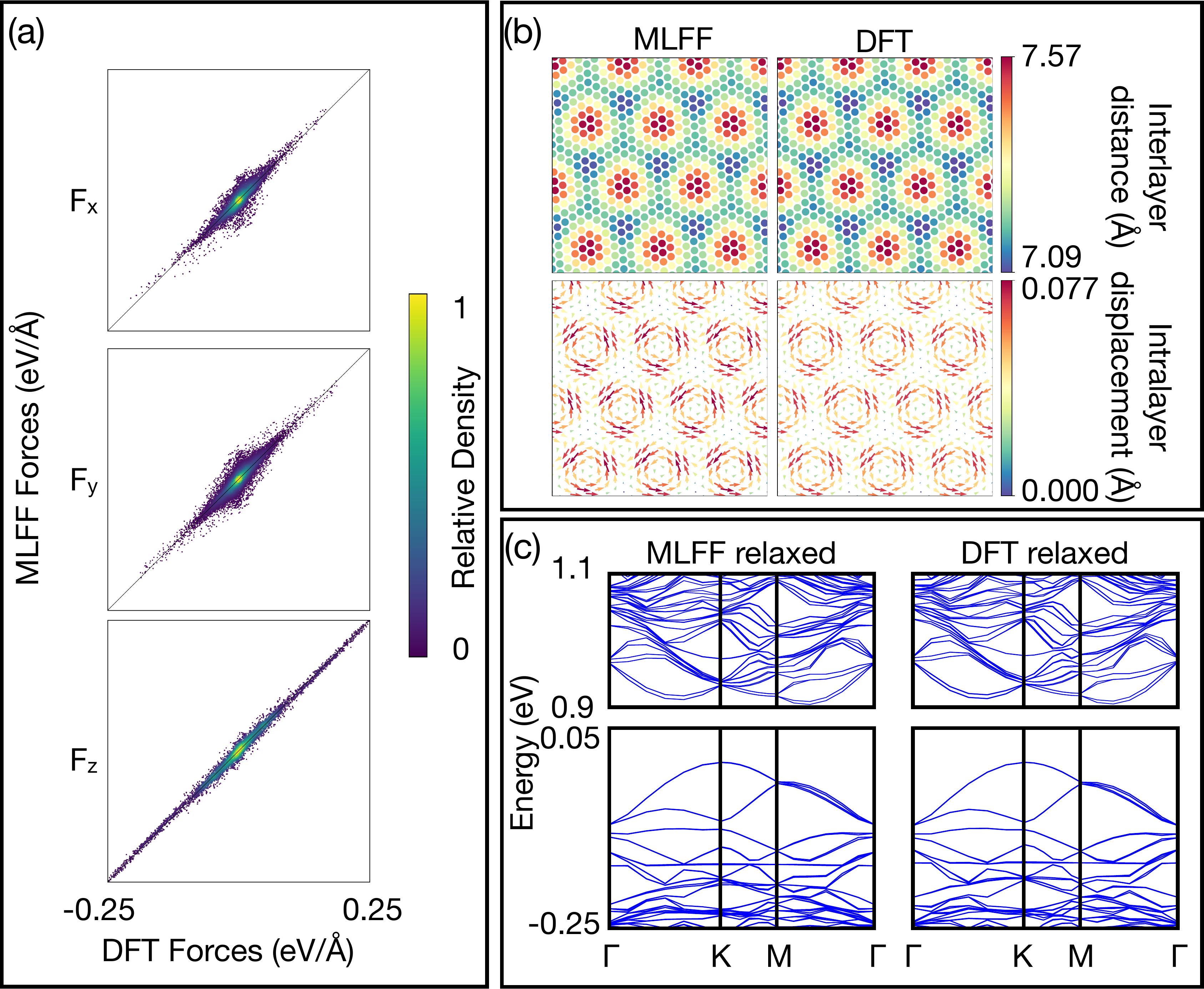}
    \caption{Evaluation of 7.34$^{\circ}$ AB MoTe$_2$. (a), Comparison between MLFF-predicted forces and DFT-calculated forces in test set. The x-axis shows the DFT-calculated force, and the y-axis shows the MLFF-calculated force. (b), Relaxation pattern of MLFF-relaxed structure and DFT-relaxed structure. Figs in first row are MLFF-relaxed structure; Figs in second row are DFT-relaxed structure.
    (c), Band structure of MLFF-relaxed structure and DFT-relaxed structure.}
    \label{fig:AB_MoTe2}
\end{figure}

\begin{figure}
    \centering
    \includegraphics[width=1\linewidth]{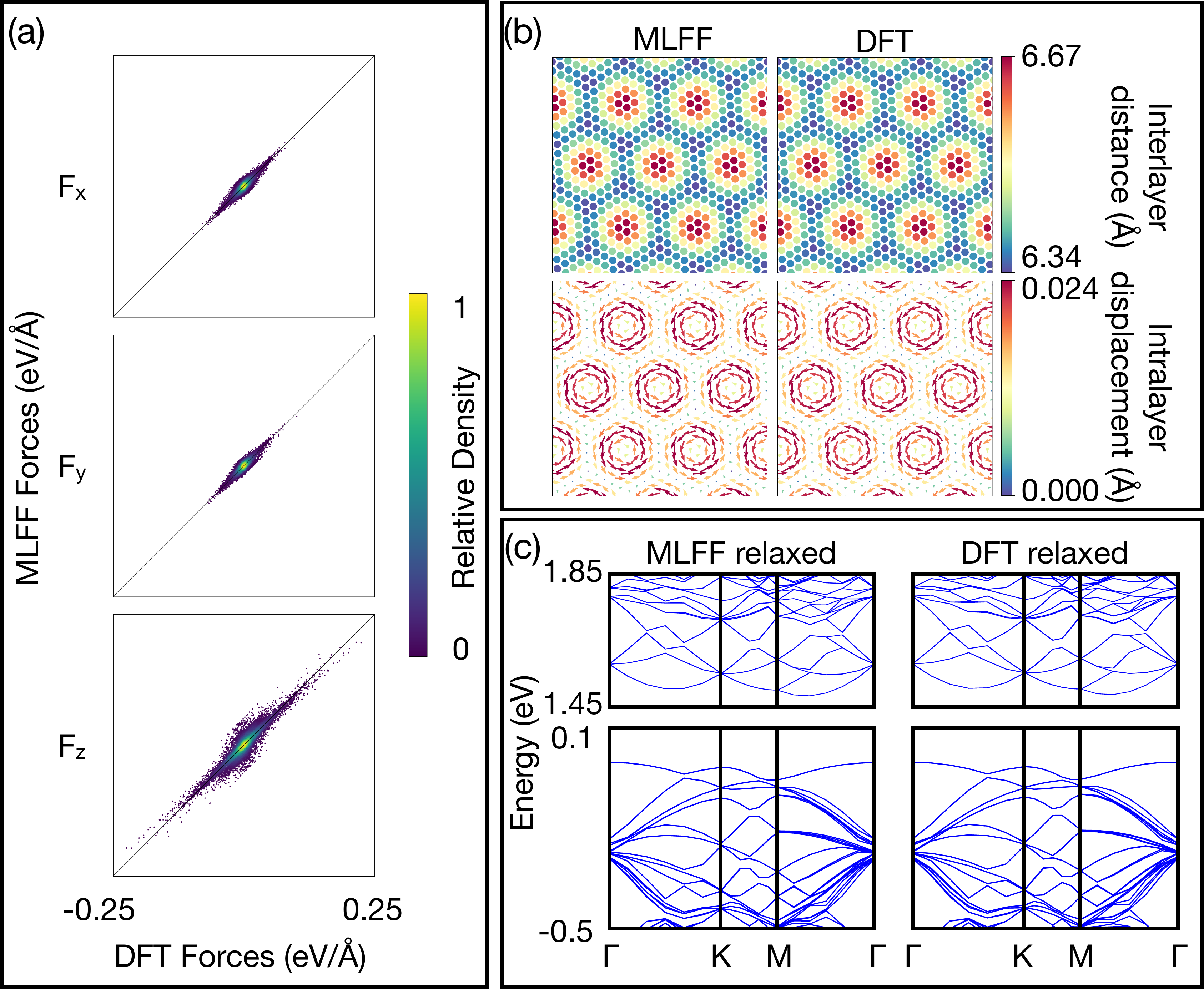}
    \caption{Evaluation of 7.34$^{\circ}$ AA WS$_2$. (a), Comparison between MLFF-predicted forces and DFT-calculated forces in test set. The x-axis shows the DFT-calculated force, and the y-axis shows the MLFF-calculated force. (b), Relaxation pattern of MLFF-relaxed structure and DFT-relaxed structure. Figs in first row are MLFF-relaxed structure; Figs in second row are DFT-relaxed structure.
    (c), Band structure of MLFF-relaxed structure and DFT-relaxed structure.}
    \label{fig:AA_WS2}
\end{figure}

 \begin{figure}
    \centering
    \includegraphics[width=1\linewidth]{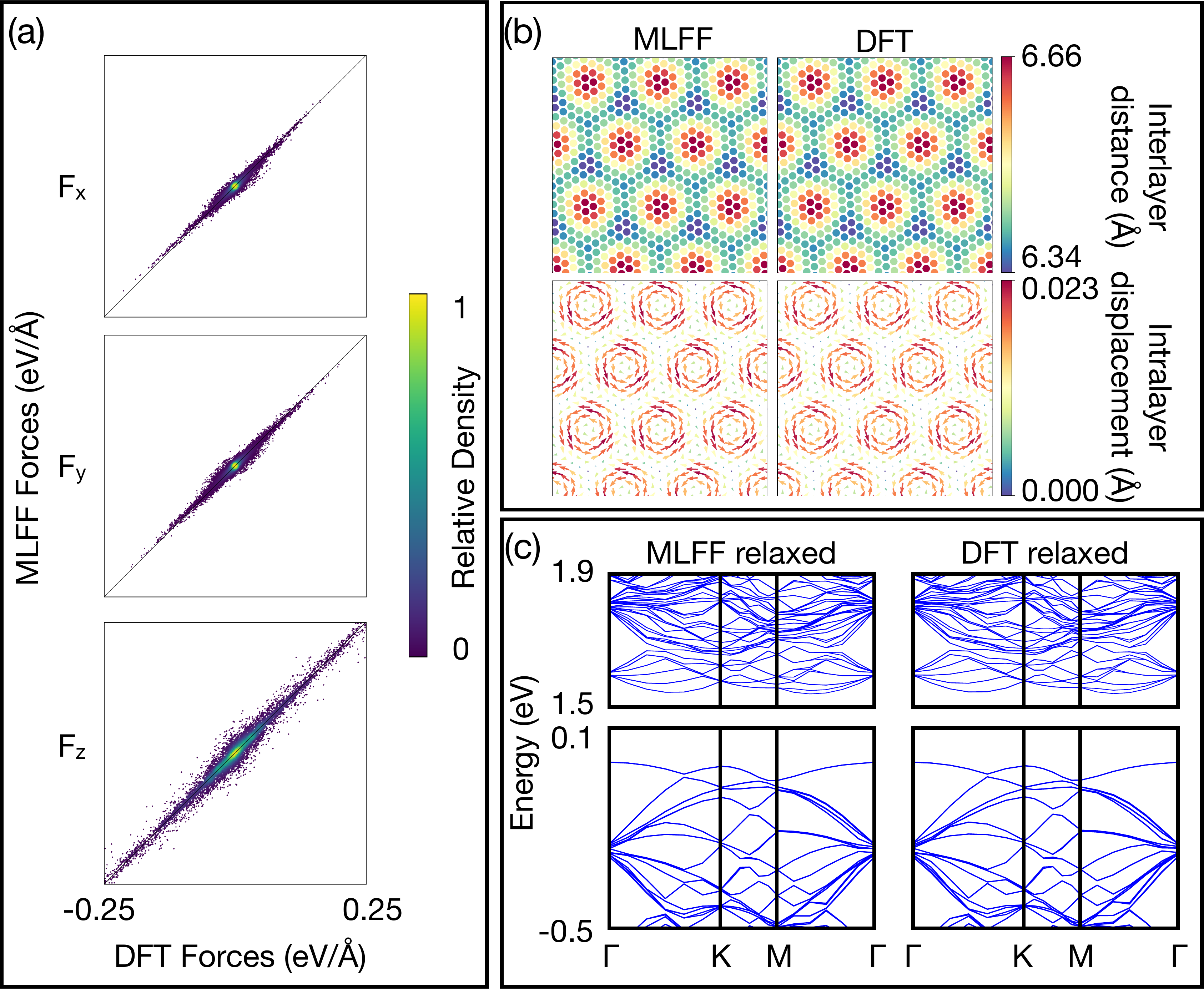}
    \caption{Evaluation of 7.34$^{\circ}$ AB WS$_2$. (a), Comparison between MLFF-predicted forces and DFT-calculated forces in test set. The x-axis shows the DFT-calculated force, and the y-axis shows the MLFF-calculated force. (b), Relaxation pattern of MLFF-relaxed structure and DFT-relaxed structure. Figs in first row are MLFF-relaxed structure; Figs in second row are DFT-relaxed structure.
    (c), Band structure of MLFF-relaxed structure and DFT-relaxed structure.}
    \label{fig:AB_WS2}
\end{figure}

\begin{figure}
    \centering
    \includegraphics[width=1\linewidth]{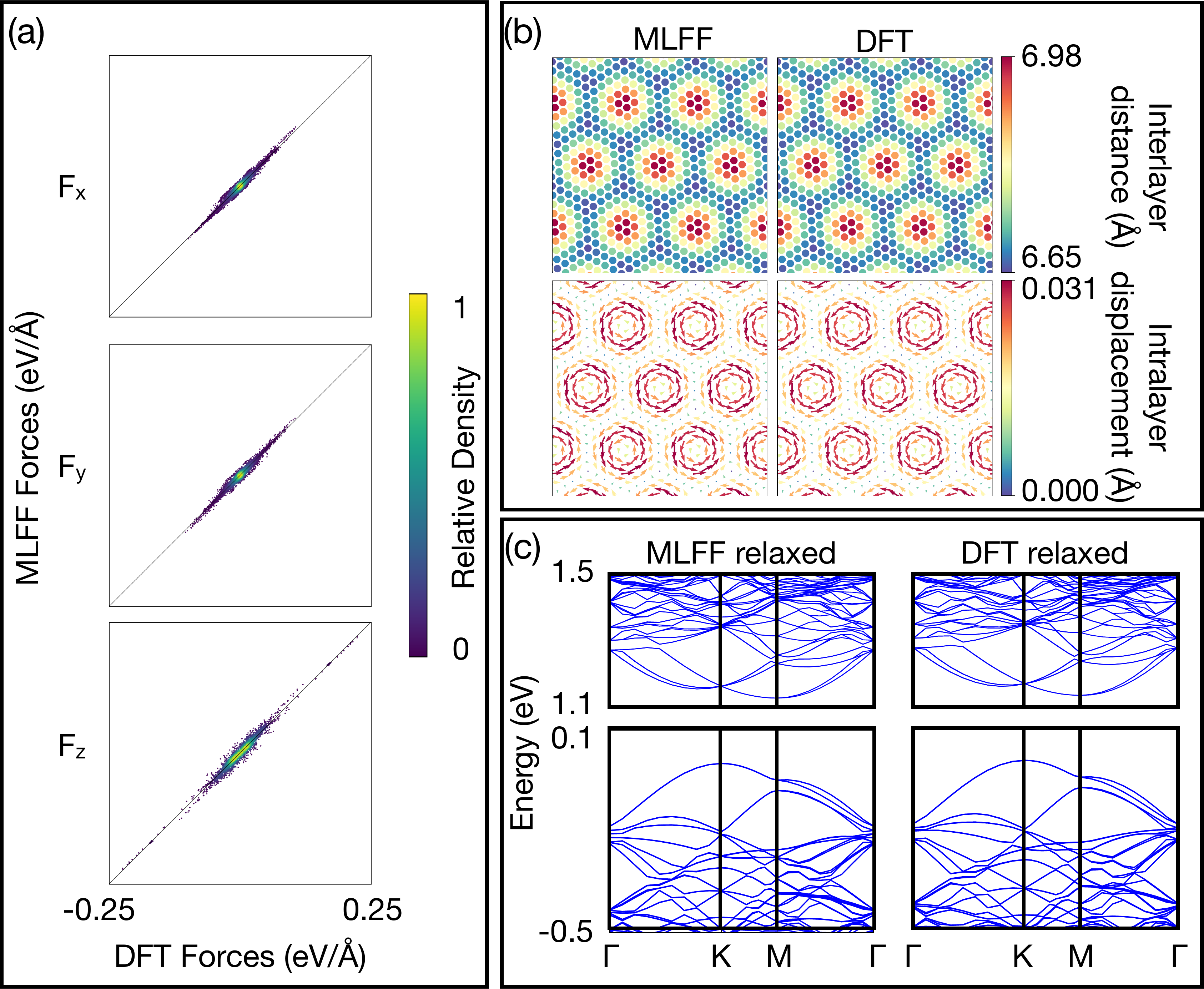}
    \caption{Evaluation of 7.34$^{\circ}$ AA WSe$_2$. (a), Comparison between MLFF-predicted forces and DFT-calculated forces in test set. The x-axis shows the DFT-calculated force, and the y-axis shows the MLFF-calculated force. (b), Relaxation pattern of MLFF-relaxed structure and DFT-relaxed structure. Figs in first row are MLFF-relaxed structure; Figs in second row are DFT-relaxed structure.
    (c), Band structure of MLFF-relaxed structure and DFT-relaxed structure.}
    \label{fig:AA_WSe2}
\end{figure}

 \begin{figure}
    \centering
    \includegraphics[width=1\linewidth]{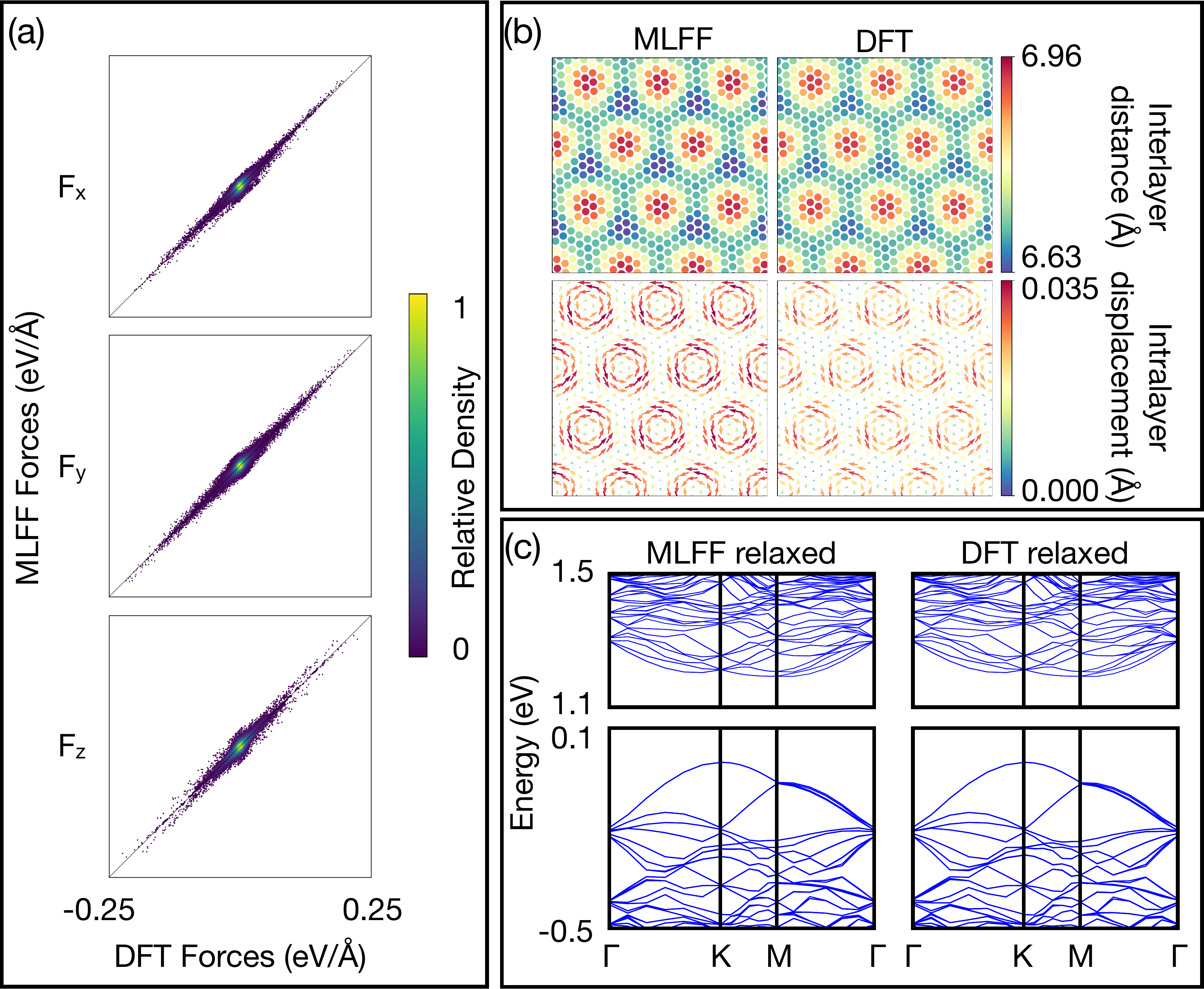}
    \caption{Evaluation of 7.34$^{\circ}$ AB WSe$_2$. (a), Comparison between MLFF-predicted forces and DFT-calculated forces in test set. The x-axis shows the DFT-calculated force, and the y-axis shows the MLFF-calculated force. (b), Relaxation pattern of MLFF-relaxed structure and DFT-relaxed structure. Figs in first row are MLFF-relaxed structure; Figs in second row are DFT-relaxed structure.
    (c), Band structure of MLFF-relaxed structure and DFT-relaxed structure.}
    \label{fig:AB_WSe2}
\end{figure}

\begin{figure}
    \centering
    \includegraphics[width=1\linewidth]{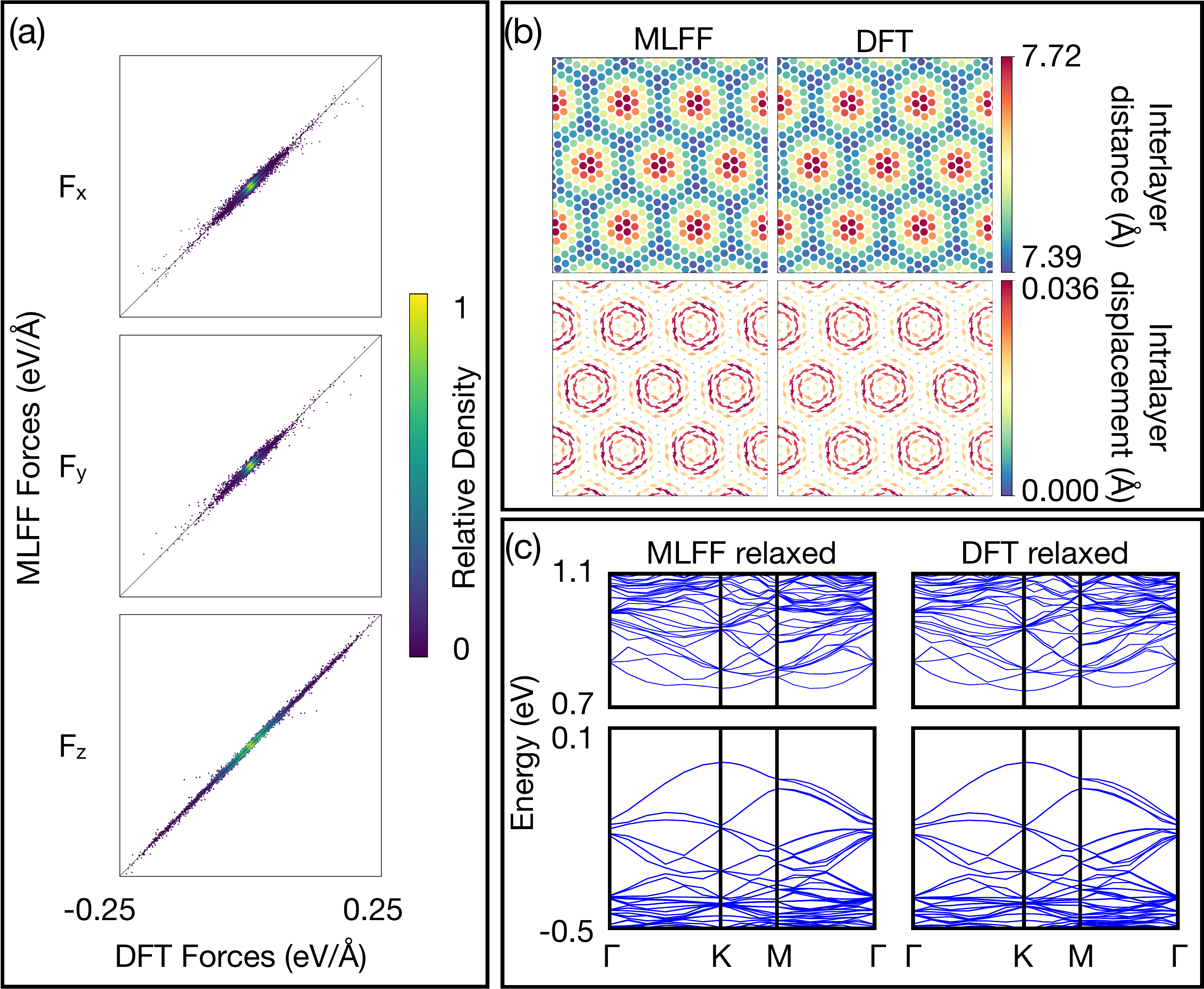}
    \caption{Evaluation of 7.34$^{\circ}$ AA WTe$_2$. (a), Comparison between MLFF-predicted forces and DFT-calculated forces in test set. The x-axis shows the DFT-calculated force, and the y-axis shows the MLFF-calculated force. (b), Relaxation pattern of MLFF-relaxed structure and DFT-relaxed structure. Figs in first row are MLFF-relaxed structure; Figs in second row are DFT-relaxed structure.
    (c), Band structure of MLFF-relaxed structure and DFT-relaxed structure.}
    \label{fig:AA_WTe2}
\end{figure}

 \begin{figure}
    \centering
    \includegraphics[width=1\linewidth]{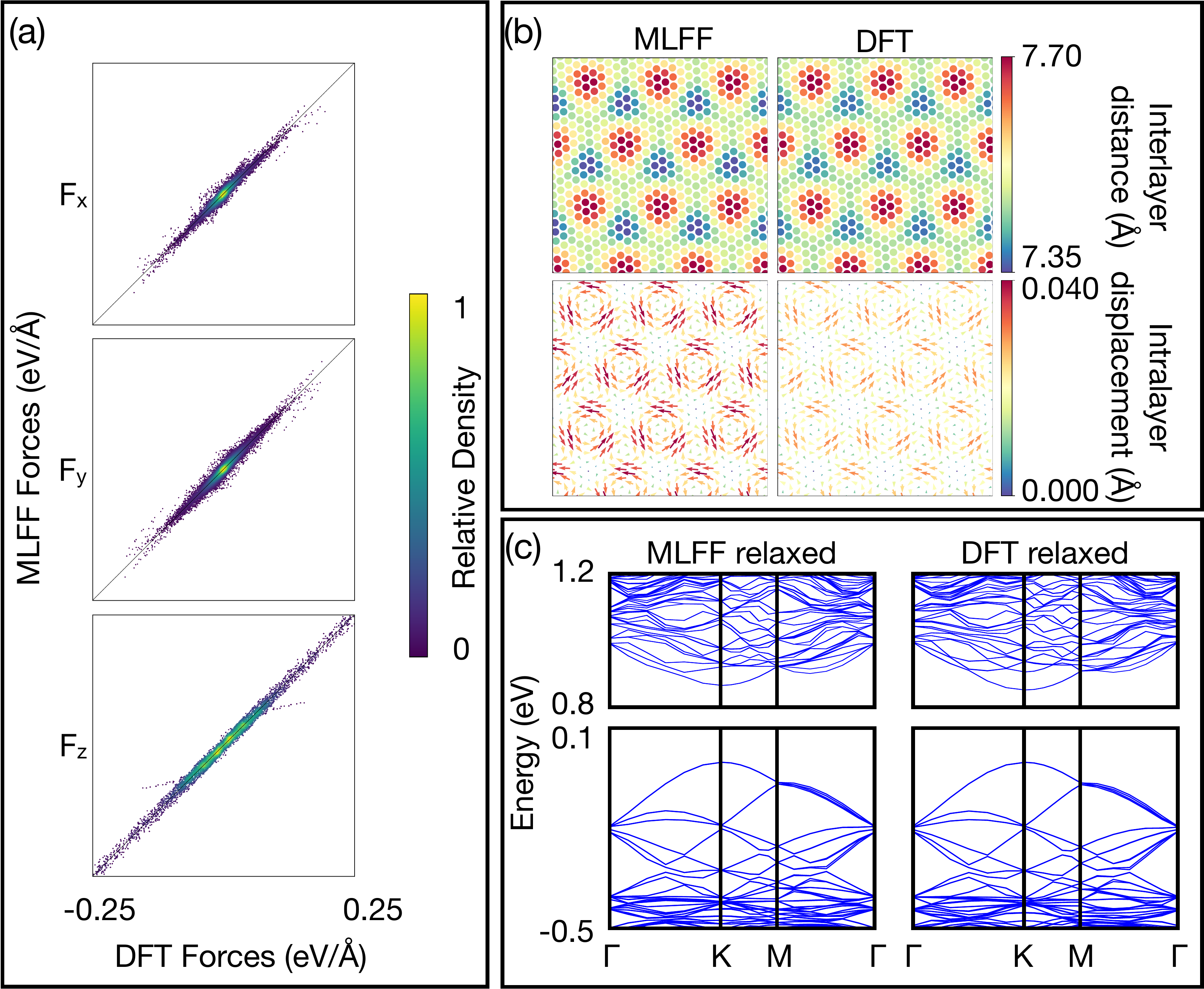}
    \caption{Evaluation of 7.34$^{\circ}$ AB WTe$_2$. (a), Comparison between MLFF-predicted forces and DFT-calculated forces in test set. The x-axis shows the DFT-calculated force, and the y-axis shows the MLFF-calculated force. (b), Relaxation pattern of MLFF-relaxed structure and DFT-relaxed structure. Figs in first row are MLFF-relaxed structure; Figs in second row are DFT-relaxed structure.
    (c), Band structure of MLFF-relaxed structure and DFT-relaxed structure.}
    \label{fig:AB_WTe2}
\end{figure}

 \begin{figure*}
    \centering
    \includegraphics[width=1\linewidth]{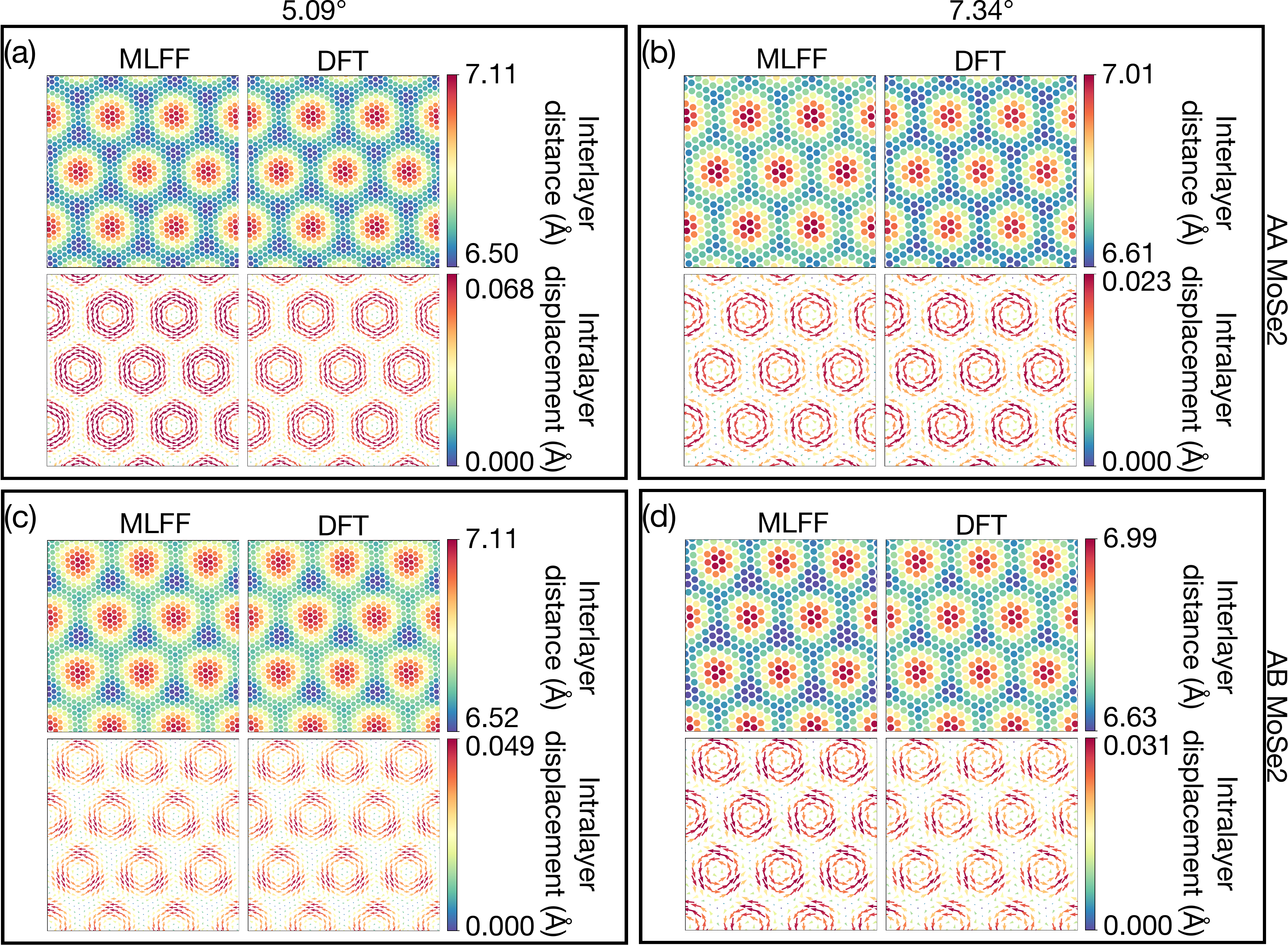}
    \caption{Comparison between 5.09$^{\circ}$ and 7.34$^{\circ}$ MoSe$_2$. 
    (a), Relaxation pattern of 5.09$^{\circ}$ MLFF-relaxed structure and DFT-relaxed AA MoSe2.
    (b), Relaxation pattern of 7.34$^{\circ}$ MLFF-relaxed structure and DFT-relaxed AA MoSe2.
    (c), Relaxation pattern of 5.09$^{\circ}$ MLFF-relaxed structure and DFT-relaxed AB MoSe2.
    (d), Relaxation pattern of 7.34$^{\circ}$ MLFF-relaxed structure and DFT-relaxed AB MoSe2.}
    \label{fig:MoSe2_5.09}
\end{figure*}
\clearpage
\bibliography{refs}% Produces the bibliography via BibTeX.

\end{document}